\documentclass[fleqn,usenatbib]{mnras}

\usepackage{newtxtext,newtxmath}

\usepackage[T1]{fontenc}
\usepackage{ae,aecompl}
\usepackage{cite}

\usepackage{graphicx}	
\usepackage{amsmath}	
\usepackage{longtable}
\usepackage{booktabs}
\usepackage{geometry,color,graphicx,float}

\title[A thermophysical and dynamical study of the Hildas (1162) Larissa and (1911) Schubart]{A thermophysical and dynamical study of the Hildas (1162) Larissa and (1911) Schubart}

\author[C. F. Chavez et al.]{
Cristian F. Chavez$^{1,2},$\thanks{E-mail: cchavez.astro.usq@gmail.com}
T. G. M\"uller$^{3},$
J. P. Marshall$^{4,1},$ 
J. Horner$^{1},$
H. Drass$^{2,1},$
B. Carter$^{1}$
\\ 
$^{1}$Centre for Astrophysics, University of Southern Queensland, West St, Darling Heights, Toowoomba, QLD 4350, Australia\\
$^{2}$Centre for Astro-Engineering, Pontificia Universidad Cat\'olica de Chile, Av. Vicuña Mackenna 4860, Santiago, Chile\\
$^{3}$Max-Planck-Institut f\"ur Extraterrestrische Physik, Giessenbachstraße 1, 85748 Garching, Germany\\
$^{4}$Academia Sinica Institute of Astronomy and Astrophysics, AS/NTU Astronomy-Mathematics Building, No.1, Sect. 4, Roosevelt Rd,\\ Taipei 10617, Taiwan\\
}

\date{Accepted XXX. Received YYY; in original form ZZZ}

\pubyear{2020}

\begin{document}
\label{firstpage}
\pagerange{\pageref{firstpage}--\pageref{lastpage}}
\maketitle
 
\begin{abstract}
The Hilda asteroids are among the least studied populations in the asteroid belt, despite their potential importance as markers of Jupiter's migration in the early Solar system. We present new mid-infrared observations of two notable Hildas, (1162) Larissa and (1911) Schubart, obtained using the Faint Object infraRed CAmera for the SOFIA Telescope (FORCAST), and use these to characterise their thermal inertia and physical properties. For (1162) Larissa, we obtain an effective diameter of
\textcolor{black}{46.5$^{+2.3}_{-1.7}$~km, an albedo of 0.12~$\pm$~0.02, and a thermal inertia of 15$^{+10}_{-8}$ Jm$^{-2}$s$^{1/2}$K$^{-1}$. In addition, our Larissa thermal measurements are well matched with an ellipsoidal shape with an axis ratio a/b=1.2 for the most-likely spin properties. Our modelling of (1911) Schubart is not as refined, but the thermal data point towards a high-obliquity spin-pole, with a best-fit a/b=1.3 ellipsoidal shape. This spin-shape solution is yielding a diameter of 72$^{+3}_{-4}$ km, an albedo of 0.039$\pm$~0.02, and a thermal inertia below 30 Jm$^{-2}$s$^{1/2}$K$^{-1}$ (or 10$^{+20}_{-5}$ Jm$^{-2}$s$^{1/2}$K$^{-1}$).} As with (1162) Larissa, our results suggest that (1911) Schubart is aspherical, and likely elongated in shape. Detailed dynamical simulations of the two Hildas reveal that both exhibit strong dynamical stability, behaviour that suggests that they are primordial, rather than captured objects. The differences in their albedos, along with their divergent taxonomical classification, suggests that despite their common origin, the two have experienced markedly different histories. \\ 
\end{abstract}




\section{Introduction}

Over the past few decades, there has been a growing consensus amongst astronomers and planetary scientists that the Solar system's youth was a chaotic place. Rather than being the result of a sedate, in-situ formation, leading models now suggest that the terrestrial planets were shaped by cataclysmic collisions \citep[e.g.][]{Earth_1,Earth_2,Merc_1,Earth_3,Merc_2}, whilst the giant planets underwent dramatic migration, potentially spanning hundreds of millions or even billions of kilometres \citep[e.g.][]{nice1,gomes,migrate2,lykawka,NT2,NT3,migrate3,migrate4}.

Much of the evidence for the chaotic evolution of the early Solar system has come, not from studies of the planets themselves, but from analysis of the vast numbers of small bodies littered throughout the system\footnote{For a detailed recent review of the various theories of Solar system's migration history, and the abundant evidence that supports them, we direct the interested reader to \citet{SSRev}, and references therein.}. The Solar system's asteroidal bodies are of particular interest in determining the true narrative of the system's past. The overwhelming majority of those objects move on orbits between those of Mars and Jupiter, within the Asteroid belt. Those asteroids are of great scientific value as natural records of the formation and evolution of our Solar system. The study of their orbits and physical properties are key research topics to understand the mechanics and evolution of the Solar system as a whole, and particularly the origin, hydration, and collisional history of the Earth {\citep[e.g.][]{veneer,Strom05,BottkeBelt,gomes,fof1,2009DH,secularsaturn,migrate4}}.

In recent years, studies of the Jovian Trojan population have brought fresh insights to the migration history of the giant planets \citep[e.g.][]{morbi05,nesv13,roig15,migrate4,nesv18b,pirani19}. The degree to which the orbits of Jupiter's Trojans are excited suggests that they must have been captured during the migration of the giant planets, rather than having formed in-situ -- and as a result, the population has been used as a means to estimate the scale, speed, direction, and chaoticity of that migration \citep[e.g.][]{morbi05,nesv13,pirani19}. 

The Jovian Trojans are not, however, the only population of asteroidal bodies trapped in mean-motion resonance with Jupiter. Located between the outer edge of the Asteroid belt and Jupiter's orbit lie the Hildas -- a swarm of objects trapped in 3:2 Mean Motion Resonance with Jupiter, at a semi-major axis of around 4 au \citep{hildas,SDSS,ages_1911_milani}. It is natural, therefore, to wonder whether this population, too, holds clues that may help us unveil the full narrative of the Solar system's early days. 

Following that logic, both \citet{roig15} and \citet{pirani19} considered the effect of Jupiter's migration on the planet's two main resonant populations. \citet{roig15} examined the ``Jumping-Jupiter'' model of Jovian migration, and find that \textit{``neither primordial Hildas, nor Trojans, survive the instability, confirming the idea that such populations must have been implanted from other sources''}. Similarly, \citet{pirani19} investigated how a rapid inward migration of Jupiter from a more distant initial orbit (beyond $\sim$~15 au) would have affected the Jovian Trojans and the Hildas. They found that such migration would result in the capture of a massive and excited Hilda population -- something that is strongly at odds with the modern observed distribution of Hildas, which move on remarkably dynamically cold orbits. Their model does, however, successfully replicate the observed asymmetry between the observed populations of the leading and trailing clouds of Jovian Trojans. These recent studies reveal the importance of the Hildas as a key test for models of planet formation, and as such, it is fair to consider the Hildas as a whole an interesting but underrated link to the Solar system's formation and evolution. 

The first Hilda asteroid to be discovered was that after which the family is named -- (153) Hilda, which was found in November 1875 by Johann Palisa{\footnote{Discovery details taken from the NASA/JPL Small-Body Database Browser at {\url{https://ssd.jpl.nasa.gov/sbdb.cgi}}, accessed on 2020 March 11.}}, some 30 years before the discovery of the first Jovian Trojans \citep[e.g.][]{achilles,patroclus,hektor}. Although the Hildas orbit markedly closer to the Sun than the Jovian Trojans, and their population is comparable to that of their more famous cousins, they remain relatively poorly studied. 

Recent studies that have focused on the physical properties of Hildas have suggested that these asteroids may have had a common origin with Jovian Trojans. Studies of their near infrared spectra \citep{spectra}, size distributions \citep{terai}, size frequency distributions \citep{yoshida}, and their optical and infrared colours (from the Sloan Digital Sky Survey; \citealt{SDSS}, and NEOWISE; \citealt{primass}), reveal that the overall properties of the two resonant populations are broadly similar. 

Such studies have revealed the presence of at least two collisional/dynamical families in the Hilda population, both of which are named for their largest members \citep[e.g.][]{broz}. To date, 111 members of the (1911) Schubart family have been identified, whilst the (153) Hilda family has at least 216 confirmed members \citep{hidas-romanishin}. It seems likely that future surveys will reveal both more members of these families, and new collisional families with different parent bodies.

Taxonomically, the detailed mapping of the asteroid belt carried out by \citet{DeMeo2013, DeMeo2014} found the Hildas and Jovian Trojans to be distinct groups of objects, separate from their classification of the main belt asteroids. The Hildas and Trojans were typically labelled as P and D-types in contrast to the S and V-types that are common throughout the main belt, a result that has been verified by later studies \citep{terai, yoshida}. In addition to their similar taxonomic classifications, the Hildas and Jovian Trojans both exhibit a clear bimodality in their spectral distribution \citep{terai,primass}, a fact that could be suggestive of a variety of source regions having contributed to the modern Hilda and Trojan populations.
 
To date, amongst the Hildas that have been classified on the basis of their spectral data, the majority of were classified under the low-albedo asteroid classifications C, P and D-type \citep{asteroids_dwarf,spectra,Szabo_K2} in the Tholen taxonomy \citep{Kaas, LCDB, Lissauer}. Of these, the D-type asteroids are undoubtedly the most numerous, followed by P-types and finally C-types \citep{Grav2012}. 

Some studies state that the general outline for the surface composition of the Hildas is a mix of organics, anhydrous silicates, opaque material and ice \citep{47,hildas_taxonomy_6,SDSS,primass}. The surfaces of the Hildas have likely experienced significant space weathering, although it should be noted that they may well be considered to be more pristine than similar objects in the main asteroid belt, as a result of their larger heliocentric distances \citep{SDSS, primass}.

Moreover, \citet{primass} raised an interesting point on current knowledge of Hildas concluding that there is a population in the Jovian Trojans not present in the Hildas,
\textcolor{black}{supporting the idea that Trojans and Hildas may be celestial objects of different origins. Indeed, the stark contrast in mean albedo values amongst sub-groups in the same Hildas suggests that even within them we can appreciate a kind of intruder of another source  \citep{hidas-romanishin}}. In the coming years, the Transiting Exoplanet Survey Satellite \citep[\textit{TESS};][]{TESS} will aid the identification of Hildas with different origins, by providing a wealth of high cadence light curves for these asteroids. As described by \citet{Szabo_K2}, that detailed photometric data will facilitate a direct comparison between the main belt asteroids, the Hildas, and the Jovian Trojans, allowing the similarities and differences between the populations to be studied in depth. 

In this work, we focus on two of the largest and brightest members of the Hilda population -- (1162) Larissa, and (1911) Schubart. (1162) Larissa was discovered in 1930 by German astronomer Karl Reinmuth\footnotemark[\value{footnote}], and is amongst the brightest Hildas due to its relatively high albedo (from 0.11 \citep{alilagoa} to 0.18 \citep{Grav2012} in the literature), which is markedly higher than the average of 0.055 for this group of asteroids \citep{primass}. Its taxonomy has been variously defined as M-type \citep{Grav2012}, P-type \citep{LCDB}, and most recently as X-type \citep{primass}. It does not belong to either of the currently identified collisional families within the Hilda population \citep[e.g.][]{hidas-romanishin}, and as a result, it is plausible that it might be a primordial, undisrupted body.

(1911) Schubart was discovered in 1973 by Swiss astronomer Paul Wild\footnotemark[\value{footnote}], and is the largest member of a populous collisional family (the Schubart family), containing at least 111 members \citep{broz}. The extremely low albedo calculated for it, under 0.05 in several studies \citep{Grav2012,LCDB,hidas-romanishin}, makes it one of the darkest asteroids in the Hilda group and even in the Solar system.

We describe the observations taken by the ``Faint Object infraRed CAmera for the SOFIA Telescope'' \citep[FORCAST,][]{FORCAST_0} instrument on the Stratospheric Observatory For Infrared Astronomy \citep[SOFIA,][]{first_science_obs} used in this work in Section~\ref{Sec:Observations}. The process by which flux values are extracted from the observations is then described in Section~\ref{Sec:Analysis}. In Section~\ref{Sec:Thermophysical}, we perform a detailed thermophysical analysis of (1162) Larissa and (1911) Schubart, before investigating the orbital stability of the two asteroids in Section~\ref{Sec:Dynamics}. In Section~\ref{Sec:Conclusions}, we discuss our results and present our conclusions.\\

\section{Observations}
\label{Sec:Observations}
We obtained SOFIA/FORCAST observations of the Hildas (1162) Larissa and (1911) Schubart, which we supplement with archival mid- and far-infrared data taken by \textit{IRAS} \citep{tedesco1986}, \textit{AKARI} \citep{usui} and \textit{WISE} \citep{mainzer} projects. The SOFIA images collected for this research were part of a series of studies about small bodies and Solar system formation that were performed as part of the so-called ``Basic Science'' observation flights, under the proposal ID 82$\_$0004, mission ID FO64, PI Dr. Thomas M\"uller, carried out during 2011 June.

Our data comprise a series of multi-wavelength mid-infrared imaging observations taken on 2011 June 4, for (1911) Schubart, and 2011 June 8 for (1162) Larissa. FORCAST is a dual-channel infrared camera and spectrograph, able to produce images with a field of view of approximately 3.4$^{\prime}$ $\times$ 3.2$^{\prime}$, with a plate scale of 0.768$^{\prime\prime}$. One channel is the short-wavelength camera (SWC), which operates at 5 to 26 $\mu$m, whilst the other, the long-wavelength camera (LWC), operates at 26 to 40 $\mu$m; both channels can be used either individually or simultaneously \citep{data_red_sofia}.

In order to take advantage of the dual channel nature of FORCAST, two observations were taken per target, in couplet using first the 11.1 and 34.8 $\mu$m and then the 19.7 and 31.5 $\mu$m filters, under the symmetrical two-position chop-nod mode (C2N), to minimise the impact of a variety of sources of noise, including time variable sky backgrounds and the thermal emission from the telescope itself. That mode is recommended for observations of point-like sources such as (1162) Larissa and (1911) Schubart, as described in \citet{first_science_obs}.

The infrared images used in this work are level 3 coadded data products, which means that they were telluric and flux corrected by SOFIA team in units of Jy/pix \citep{data_red_sofia}. The files were stored and then downloaded from the \textsc{SOFIA Airborne Observatory Public Archive}\footnote{\url{https://www.sofia.usra.edu/science/data/science-archive/} Accessed on 2019 October 10}.

To complement our new data, we searched for archival observations of our two targets. Fortunately, both (1162) Larissa and (1911) Schubart have been observed multiple times at thermal infrared wavelengths by other surveys, so that in addition to the 12 SOFIA FORCAST images of (1162) Larissa, for the thermophysical analysis we used one observation from \textit{IRAS} (at 25 $\mu$m), four from \textit{AKARI} (at 18$\mu$m), and 42 from \textit{WISE} (21 images at 11.1 $\mu$m and 21 at 22.64 $\mu$m).

Moreover, for (1911) Schubart\textcolor{black}{, in addition to the 23 SOFIA FORCAST images,} we used 14 observations from \textit{IRAS} (four at 12 $\mu$m, four at 25 $\mu$m, four at 60 $\mu$m and two at 100 $\mu$m), eight from \textit{AKARI} (four at 9 $\mu$m and four at 18 $\mu$m) and also 39 from \textit{WISE} (20 at 11.1 $\mu$m and 19 at 22.64 $\mu$m). All the data for both asteroids were taken from the SBNAF Infrared Database \citep{2020_DB} at VizieR Online Data Catalog{\footnote{VizieR Online Data Catalog: SBNAF Infrared Database Website at {\url{http://vizier.u-strasbg.fr/viz-bin/VizieR?-source=J/A+A/635/A54}}, accessed most recently on 2020 July 14. Earlier in our analysis, we obtained data from the Konkoly Observatory Website at {\url{https://konkoly.hu/database.shtml}}, accessed on 2019 December 17, which was used for our preliminary work, but has since been superseded by the updated VizieR catalog.}}. Full details on the archival images used are given in Tables~\ref{tab:table1} and \ref{tab:table2}.

\begin{center}
\begin{table*}
{\small
\hfill{}
\begin{tabular}{ccccccccc}
\hline
\textbf{Date [JD]}&\textbf{Wavelength [$\mu$m]}
&\textbf{Flux density [Jy]}&\textbf{$r_{\rm Helio}$ [au]}&\textbf{$\Delta$ [au]}&\textbf{$\alpha$ [$^{\circ}$]}&\textbf{Telescope/Instrument} \\
\hline
2445613.4677 &25.00 &0.538~$\pm$~0.141 &4.38557 &4.07305 &12.90 &IRAS\\

2453982.9395 &18.00 &0.221~$\pm$~0.033 &4.36022 &4.25199 &13.37 &AKARI-IRC\\

2453983.0083 &18.00 &0.299~$\pm$~0.038 &4.36023 &4.25095 &13.37 &AKARI-IRC\\

2454154.3825 &18.00 &0.298~$\pm$~0.038 &4.37176 &4.26789 &-13.08 &AKARI-IRC\\

2454154.4514 &18.00 &0.327~$\pm$~0.040 &4.37175 &4.26896 &-13.08 &AKARI-IRC\\

2455225.2251 &11.56 &0.223~$\pm$~0.011 &3.61143 &3.49250 &15.82 &WISE\\

2455225.2251 &22.09 &0.654~$\pm$~0.046 &3.61143 &3.49250 &15.82 &WISE\\

2455225.3576 &11.56 &0.180~$\pm$~0.009 &3.61134 &3.49043 &15.82 &WISE\\

2455225.3576 &22.09 &0.624~$\pm$~0.040 &3.61134 &3.49043 &15.82 &WISE\\

2455225.4899 &11.56 &0.226~$\pm$~0.011 &3.61124 &3.48837 &15.83 &WISE\\

2455225.4899 &22.09 &0.708~$\pm$~0.044 &3.61124 &3.48837 &15.83 &WISE\\

2455225.6222 &11.56 &0.183~$\pm$~0.009 &3.61114 &3.48630 &15.83 &WISE\\

2455225.6222 &22.09 &0.601~$\pm$~0.038 &3.61114 &3.48630 &15.83 &WISE\\

2455225.6884 &11.56 &0.168~$\pm$~0.008 &3.61109 &3.48527 &15.83 &WISE\\

2455225.6884 &22.09 &0.554~$\pm$~0.033 &3.61109 &3.48527 &15.83 &WISE\\

2455225.7545 &11.56 &0.231~$\pm$~0.011 &3.61105 &3.48424 &15.83 &WISE\\

2455225.7545 &22.09 &0.725~$\pm$~0.045 &3.61105 &3.48424 &15.83 &WISE\\

2455225.9530 &11.56 &0.162~$\pm$~0.008 &3.61090 &3.48115 &15.83 &WISE\\

2455225.9530 &22.09 &0.525~$\pm$~0.033 &3.61090 &3.48115 &15.83 &WISE\\

2455226.0192 &11.56 &0.232~$\pm$~0.011 &3.61085 &3.48011 &15.83 &WISE\\

2455226.0192 &22.09 &0.733~$\pm$~0.048 &3.61085 &3.48011 &15.83 &WISE\\

2455226.0854 &11.56 &0.160~$\pm$~0.008 &3.61081 &3.47908 &15.83 &WISE\\

2455226.0854 &22.09 &0.508~$\pm$~0.033 &3.61081 &3.47908 &15.83 &WISE\\

2455226.2177 &11.56 &0.163~$\pm$~0.008 &3.61071 &3.47702 &15.83 &WISE\\

2455226.2177 &22.09 &0.526~$\pm$~0.032 &3.61071 &3.47702 &15.83 &WISE\\

2455226.3500 &11.56 &0.172~$\pm$~0.008 &3.61061 &3.47495 &15.83 &WISE\\

2455226.3500 &22.09 &0.532~$\pm$~0.034 &3.61061 &3.47495 &15.83 &WISE\\

2455399.6451 &11.56 &0.236~$\pm$~0.011 &3.51466 &3.28306 &-16.75 &WISE\\

2455399.6451 &22.09 &0.671~$\pm$~0.042 &3.51466 &3.28306 &-16.75 &WISE\\

2455399.7774 &11.56 &0.230~$\pm$~0.011 &3.51461 &3.28488 &-16.75 &WISE\\

2455399.7774 &22.09 &0.674~$\pm$~0.044 &3.51461 &3.28488 &-16.75 &WISE\\

2455399.9097 &11.56 &0.238~$\pm$~0.011 &3.51456 &3.28670 &-16.75 &WISE\\

2455399.9097 &22.09 &0.749~$\pm$~0.045 &3.51456 &3.28670 &-16.75 &WISE\\

2455400.0420 &11.56 &0.229~$\pm$~0.011 &3.51452 &3.28852 &-16.75 &WISE\\

2455400.0420 &22.09 &0.701~$\pm$~0.048 &3.51452 &3.28852 &-16.75 &WISE\\

2455400.1743 &11.56 &0.242~$\pm$~0.012 &3.51447 &3.29034 &-16.76 &WISE\\

2455400.1743 &22.09 &0.767~$\pm$~0.046 &3.51447 &3.29034 &-16.76 &WISE\\

2455400.4389 &11.56 &0.255~$\pm$~0.012 &3.51437 &3.29398 &-16.76 &WISE\\

2455400.4389 &22.09 &0.791~$\pm$~0.049 &3.51437 &3.29398 &-16.76 &WISE\\

2455400.5050 &11.56 &0.222~$\pm$~0.011 &3.51435 &3.29489 &-16.76 &WISE\\

2455400.5050 &22.09 &0.684~$\pm$~0.044 &3.51435 &3.29489 &-16.76 &WISE\\

2455400.6373 &11.56 &0.237~$\pm$~0.011 &3.51430 &3.29671 &-16.76 &WISE\\

2455400.6373 &22.09 &0.742~$\pm$~0.046 &3.51430 &3.29671 &-16.76 &WISE\\

2455400.7696 &11.56 &0.213~$\pm$~0.010 &3.51426 &3.29854 &-16.77 &WISE\\

2455400.7696 &22.09 &0.656~$\pm$~0.042 &3.51426 &3.29854 &-16.77 &WISE\\

2455400.9019 &11.56 &0.226~$\pm$~0.011 &3.51421 &3.30036 &-16.77 &WISE\\

2455400.9019 &22.09 &0.709~$\pm$~0.042 &3.51421 &3.30036 &-16.77 &WISE\\
\hline
\end{tabular}}
\hfill{}
\caption{Summary of the supplementary thermal observations of (1162) Larissa used in this work. The table shows the time at midpoint of the observation (in Julian Date), the reference wavelength (in $\mu$m), the monochromatic flux density (colour corrected in-band flux density) with its absolute error, the heliocentric distance ($r_{\rm Helio}$), the observatory-centric distance ($\Delta$; in au), the phase angle ($\alpha$), and the spacecraft telescope used for the observation.}
\label{tab:table1}
\end{table*}
\end{center}
\begin{center}
\begin{table*}
{\small
\hfill{}
\begin{tabular}{ccccccc}
\hline
\textbf{Date [JD]}&\textbf{Wavelength [$\mu$m]}
&\textbf{Flux density [Jy]}&\textbf{$r_{\rm Helio}$ [au]}&\textbf{$\Delta$ [au]}&\textbf{$\alpha$} [$^{\circ}$]&\textbf{Telescope/Instrument} \\
\hline
2445433.9806 &12.00 &1.623~$\pm$~0.286 &3.36036 &3.13997 &-17.30 &IRAS\\

2445433.9806 &25.00 &3.073~$\pm$~0.591 &3.36036 &3.13997 &-17.30 &IRAS\\

2445433.9806 &60.00 &1.559~$\pm$~0.364 &3.36036 &3.13997 &-17.30 &IRAS\\

2445433.9806 &100.00 &1.422~$\pm$~0.367 &3.36036 &3.13997 &-17.30 &IRAS\\

2445434.0522 &12.00 &1.483~$\pm$~0.291 &3.36039 &3.14101 &-17.30 &IRAS\\

2445434.0522 &25.00 &2.723~$\pm$~0.543 &3.36039 &3.14101 &-17.30 &IRAS\\

2445434.0522 &60.00 &1.615~$\pm$~0.419 &3.36039 &3.14101 &-17.30 &IRAS\\

2445434.0522 &100.00 &1.007~$\pm$~0.248 &3.36039 &3.14101 &-17.30 &IRAS\\

2445441.8569 &12.00 &1.497~$\pm$~0.256 &3.36382 &3.25466 &-17.35 &IRAS\\

2445441.8569 &25.00 &3.402~$\pm$~0.604 &3.36382 &3.25466 &-17.35 &IRAS\\

2445441.8569 &60.00 &1.704~$\pm$~0.438 &3.36382 &3.25466 &-17.35 &IRAS\\

2445441.9287 &12.00 &1.651~$\pm$~0.288 &3.36386 &3.25571 &-17.35 &IRAS\\

2445441.9287 &25.00 &3.173~$\pm$~0.608 &3.36386 &3.25571 &-17.35 &IRAS\\

2445441.9287 &60.00 &1.440~$\pm$~0.332 &3.36386 &3.25571 &-17.35 &IRAS\\

2454042.7960 &18.00 &2.377~$\pm$~0.199 &3.32468 &3.18305 &17.36 &AKARI-IRC\\

2454042.8649 &18.00 &2.417~$\pm$~0.202 &3.32469 &3.18208 &17.36 &AKARI-IRC\\

2454043.2094 &9.00 &0.507~$\pm$~0.033 &3.32472 &3.17719 &17.36 &AKARI-IRC\\

2454043.2783 &9.00 &0.442~$\pm$~0.029 &3.32473 &3.17621 &17.36 &AKARI-IRC\\

2454218.3471 &9.00 &0.468~$\pm$~0.031 &3.40787 &3.25934 &-17.18 &AKARI-IRC\\

2454218.4161 &9.00 &0.435~$\pm$~0.029 &3.40792 &3.26037 &-17.18 &AKARI-IRC\\

2454218.7612 &18.00 &2.044~$\pm$~0.172 &3.40821 &3.26552 &-17.18 &AKARI-IRC\\

2454218.8303 &18.00 &2.241~$\pm$~0.189 &3.40826 &3.26655 &-17.18 &AKARI-IRC\\

2455290.9921 &11.56 &0.156~$\pm$~0.009 &4.60277 &4.51960 &12.55 &WISE\\

2455290.9921 &22.09 &0.707~$\pm$~0.045 &4.60277 &4.51960 &12.55 &WISE\\

2455291.2567 &11.56 &0.174~$\pm$~0.010 &4.60287 &4.51558 &12.55 &WISE\\

2455291.2567 &22.09 &0.749~$\pm$~0.044 &4.60287 &4.51558 &12.55 &WISE\\

2455291.3228 &11.56 &0.179~$\pm$~0.010 &4.60290 &4.51457 &12.55 &WISE\\

2455291.3228 &22.09 &0.672~$\pm$~0.044 &4.60290 &4.51457 &12.55 &WISE\\

2455291.3890 &11.56 &0.177~$\pm$~0.010 &4.60293 &4.51357 &12.55 &WISE\\

2455291.3890 &22.09 &0.713~$\pm$~0.047 &4.60293 &4.51357 &12.55 &WISE\\

2455291.5213 &11.56 &0.161~$\pm$~0.009 &4.60298 &4.51156 &12.55 &WISE\\

2455291.5213 &22.09 &0.719~$\pm$~0.045 &4.60298 &4.51156 &12.55 &WISE\\

2455291.5874 &11.56 &0.178~$\pm$~0.010 &4.60300 &4.51055 &12.55 &WISE\\

2455291.5874 &22.09 &0.760~$\pm$~0.046 &4.60300 &4.51055 &12.55 &WISE\\

2455291.5875 &11.56 &0.175~$\pm$~0.010 &4.60300 &4.51055 &12.55 &WISE\\

2455291.5875 &22.09 &0.721~$\pm$~0.045 &4.60300 &4.51055 &12.55 &WISE\\

2455291.9844 &22.09 &0.693~$\pm$~0.045 &4.60316 &4.50452 &12.55 &WISE\\

2455293.5060 &11.56 &0.147~$\pm$~0.008 &4.60376 &4.48138 &12.56 &WISE\\

2455293.5060 &22.09 &0.671~$\pm$~0.047 &4.60376 &4.48138 &12.56 &WISE\\

2455293.5061 &11.56 &0.147~$\pm$~0.008 &4.60376 &4.48138 &12.56 &WISE\\

2455293.5061 &22.09 &0.762~$\pm$~0.052 &4.60376 &4.48138 &12.56 &WISE\\

2455293.9691 &11.56 &0.160~$\pm$~0.009 &4.60394 &4.47433 &12.56 &WISE\\

2455293.9691 &22.09 &0.719~$\pm$~0.044 &4.60394 &4.47433 &12.56 &WISE\\

2455294.0353 &22.09 &0.740~$\pm$~0.047 &4.60397 &4.47333 &12.56 &WISE\\

2455294.1014 &11.56 &0.183~$\pm$~0.010 &4.60399 &4.47232 &12.56 &WISE\\

2455294.1014 &22.09 &0.795~$\pm$~0.048 &4.60399 &4.47232 &12.56 &WISE\\

2455294.1016 &11.56 &0.198~$\pm$~0.011 &4.60399 &4.47232 &12.56 &WISE\\

2455294.1016 &22.09 &0.805~$\pm$~0.050 &4.60399 &4.47232 &12.56 &WISE\\

2455294.1677 &11.56 &0.187~$\pm$~0.010 &4.60402 &4.47132 &12.55 &WISE\\

2455294.1677 &22.09 &0.796~$\pm$~0.050 &4.60402 &4.47132 &12.55 &WISE\\

2455294.2339 &11.56 &0.177~$\pm$~0.010 &4.60404 &4.47031 &12.55 &WISE\\

2455294.2339 &22.09 &0.782~$\pm$~0.050 &4.60404 &4.47031 &12.55 &WISE\\

2455294.3662 &11.56 &0.181~$\pm$~0.010 &4.60409 &4.46830 &12.55 &WISE\\

2455294.3662 &22.09 &0.727~$\pm$~0.045 &4.60409 &4.46830 &12.55 &WISE\\

2455294.4985 &11.56 &0.165~$\pm$~0.009 &4.60415 &4.46628 &12.55 &WISE\\

2455294.4985 &22.09 &0.744~$\pm$~0.045 &4.60415 &4.46628 &12.55 &WISE\\

2455294.6308 &11.56 &0.193~$\pm$~0.011 &4.60420 &4.46427 &12.55 &WISE\\

2455294.6308 &22.09 &0.783~$\pm$~0.052 &4.60420 &4.46427 &12.55 &WISE\\

2455464.9946 &11.56 &0.165~$\pm$~0.012 &4.63763 &4.44818 &-12.45 &WISE\\

2455465.1269 &11.56 &0.169~$\pm$~0.011 &4.63763 &4.45024 &-12.45 &WISE\\

2455465.1930 &11.56 &0.162~$\pm$~0.012 &4.63763 &4.45126 &-12.45 &WISE\\
\hline
\end{tabular}}
\hfill{}
\caption{Summary of the supplementary thermal observations of (1911) Schubart used in this work. The table shows the time at the midpoint of the observation (in Julian Date), the reference wavelength (in $\mu$m), the monochromatic flux density (colour corrected in-band flux density) with its absolute error, the heliocentric distance ($r_{\rm Helio}$), the observatory-centric distance ($\Delta$, in au), the phase angle ($\alpha$), and the spacecraft telescope used for the observation.}
\label{tab:table2}
\end{table*}
\end{center}
\begin{center}
\begin{table*}
{\small
\hfill{}
\begin{tabular}{cccccccc}
\hline
\textbf{Date [JD]}&\textbf{Length of the exposure [s]}&\textbf{Wavelength [$\mu$m]}&\textbf{Flux density [Jy]}&\textbf{$r_{\rm Helio}$ [au]}& \textbf{$\Delta$ [au]}\\
\hline
2455720.9175 &65.00 &19.70 &1.849~$\pm$~0.218 &3.53485 &2.64259\\

2455720.9295 &66.72 &11.10 &0.524~$\pm$~0.040 &3.53486 &2.64250\\

2455720.9465 &66.80 &11.10 &0.458~$\pm$~0.047 &3.53486 &2.64237\\

2455720.9545 &67.80 &11.10 &0.625~$\pm$~0.051 &3.53487 &2.64231\\

2455720.9628 &63.77 &19.70 &1.665~$\pm$~0.106 &3.53487 &2.64224\\

2455720.9669 &67.23 &11.10 &0.590~$\pm$~0.056 &3.53487 &2.64221\\

2455720.9175 &65.00 &31.50 &1.630~$\pm$~0.117 &3.53485 &2.64259\\

2455720.9295 &66.72 &34.80 &1.464~$\pm$~0.110 &3.53486 &2.64250\\

2455720.9465 &66.80 &34.80 &1.241~$\pm$~0.111 &3.53486 &2.64237\\

2455720.9545 &67.80 &34.80 &1.619~$\pm$~0.134 &3.53487 &2.64231\\

2455720.9628 &63.77 &31.50 &1.508~$\pm$~0.126 &3.53487 &2.64224\\

2455720.9669 &67.23 &34.80 &1.430~$\pm$~0.118 &3.53487 &2.64221\\
\hline
\end{tabular}}
\hfill{}
\caption{Details of the observations of (1162) Larissa taken using the FORCAST instrument on the SOFIA airborne observatory on 2011 June 8, (JD 2455720).
The table shows the time at which the observations were started (in Julian Date), the length of the exposure (in seconds), the reference wavelength (in $\mu$m), the colour-corrected monochromatic flux densities with errors, the heliocentric distance of the asteroid ($r_{\rm Helio}$) and the observatory-centric distance ($\Delta$, in au). Due to the small span of time (0.05 JD), the phase angle value shows no significant change between the first and last observation, and was estimated to be 9.08$^{\circ}$.}
\label{tab:larissa_obs}
\end{table*}
\end{center}
\begin{center}
\begin{table*}
{\small
\hfill{}
\begin{tabular}{cccccccc}
\hline
\textbf{Date [JD]}&\textbf{Length of the exposure [s]}&\textbf{Wavelength [$\mu$m]}&\textbf{Flux density [Jy]}&\textbf{$r_{\rm Helio}$ [au]}& \textbf{$\Delta$ [au]}\\
\hline
2455716.9469 &63.37 &19.70 &1.084~$\pm$~0.185 &4.56625 &4.13320\\

2455716.9484 &64.78 &19.70 &1.182~$\pm$~0.174 &4.56625 &4.13318\\

2455716.9500 &63.41 &19.70 &1.028~$\pm$~0.148 &4.56625 &4.13315\\

2455716.9540 &63.54 &19.70 &1.063~$\pm$~0.156 &4.56625 &4.13309\\

2455716.9578 &64.42 &19.70 &1.099~$\pm$~0.156 &4.56625 &4.13303\\

2455716.9646 &61.34 &19.70 &1.139~$\pm$~0.192 &4.56624 &4.13293\\

\textit{2455716.9659} &\textit{61.36} &\textit{11.10} &- &- &-\\

\textit{2455716.9699} &\textit{61.42} &\textit{11.10} &- &- &-\\

\textit{2455716.9736} &\textit{61.14} &\textit{11.10} &- &- &-\\

\textit{2455716.9752} &\textit{61.01} &\textit{11.10} &- &- &-\\

\textit{2455716.9789} &\textit{61.45} &\textit{11.10} &- &- &-\\

\textit{2455716.9825} &\textit{60.80} &\textit{11.10} &- &- &-\\

\textit{2455716.9469} &\textit{63.37} &\textit{31.50} &- &- &-\\

\textit{2455716.9484} &\textit{64.78} &\textit{31.50} &- &- &-\\

2455716.9500 &63.41 &31.50 &1.112~$\pm$~0.175 &4.56625 &4.13315\\

2455716.9540 &63.54 &31.50 &1.109~$\pm$~0.169 &4.56625 &4.13309\\

2455716.9578 &64.42 &31.50 &1.132~$\pm$~0.162 &4.56625 &4.13303\\

\textit{2455716.9659} &\textit{61.36} &\textit{34.80} &- &- &-\\

\textit{2455716.9699} &\textit{61.42} &\textit{34.80} &- &- &-\\

\textit{2455716.9736} &\textit{61.14} &\textit{34.80} &- &- &-\\

2455716.9752 &61.01 &34.80 &0.879~$\pm$~0.138 &4.56624 &4.13277\\

2455716.9789 &61.45 &34.80 &1.011~$\pm$~0.173 &4.56624 &4.13271\\

\textit{2455716.9825} &\textit{60.80} &\textit{34.80} &- &- &-\\
\hline
\end{tabular}
\hfill{}
\caption{Details of the observations of (1911) Schubart taken using the FORCAST instrument on the SOFIA airborne observatory on 2011 June 4, (JD 2455716). 
The table shows the time at which the observations were started (in Julian Date), the length of the exposure (in seconds), the reference wavelength (in $\mu$m), the colour-corrected monochromatic flux densities with errors, the heliocentric distance of the asteroid ($r_{\rm Helio}$) and the observatory-centric distance ($\Delta$, in au). The images whose values are given in italics were of too poor quality to be used further in our analysis -- the asteroid was not clearly visible in those images, and any analysis therefore yielded results with very low signal-to-noise. Due to the small span of time (0.03 JD), the phase angle value shows no significant change between the first and last observation, and was estimated to be 12.12$^{\circ}$.}
 \label{tab:schubart_obs}}
\end{table*}
\end{center}

\section{Analysis of SOFIA data}
\label{Sec:Analysis}
In order to perform a thorough thermophysical analysis of the Hildas (1162) Larissa and (1911) Schubart, we first used the images taken of the asteroids at 11.1, 19.7, 31.5 and 34.8 $\mu$m to determine the flux from the asteroids in each waveband. A visual inspection of the images was carried out using SAOImageDS9 \citep{DS9} to assess their quality. This revealed that in 12 of the \textcolor{black}{23} images of (1911) Schubart there was no clear presence of the asteroid, so those images were discarded after completing the process of checking their signal-to-noise values, and they were not analysed further. The images of (1162) Larissa were all of excellent quality and, as a result, all were used in our research.

The selected images were then analysed using a bespoke code that applied implementations of standard Python-based image analysis tools including the packages {\sc astropy} \citep{astropy} and {\sc photutils} \citep{photutils}, yielding the results shown in Tables \ref{tab:larissa_obs} and \ref{tab:schubart_obs}.

Although the FORCAST Photometry Recipe reference document\footnote{SOFIA FORCAST Photometry Recipe at {\url{https://www.sofia.usra.edu/sites/default/files/Other/Documents/FORCAST_Photometry_Recipe.pdf}}, accessed on 2020 April 17.} recommends aperture photometry radius and annulus dimensions of 12 pixels and 15 to 25 pixels, the same document suggests the use of smaller apertures to reduce the overall uncertainty in measurement of low S/N sources. In this sense, given the quite irregular backgrounds of several of the images and the diversity amongst them, we opted to individually optimise the values for the aperture photometry radii and annuli for each image. 

\begin{figure}
	\includegraphics[width=\columnwidth]{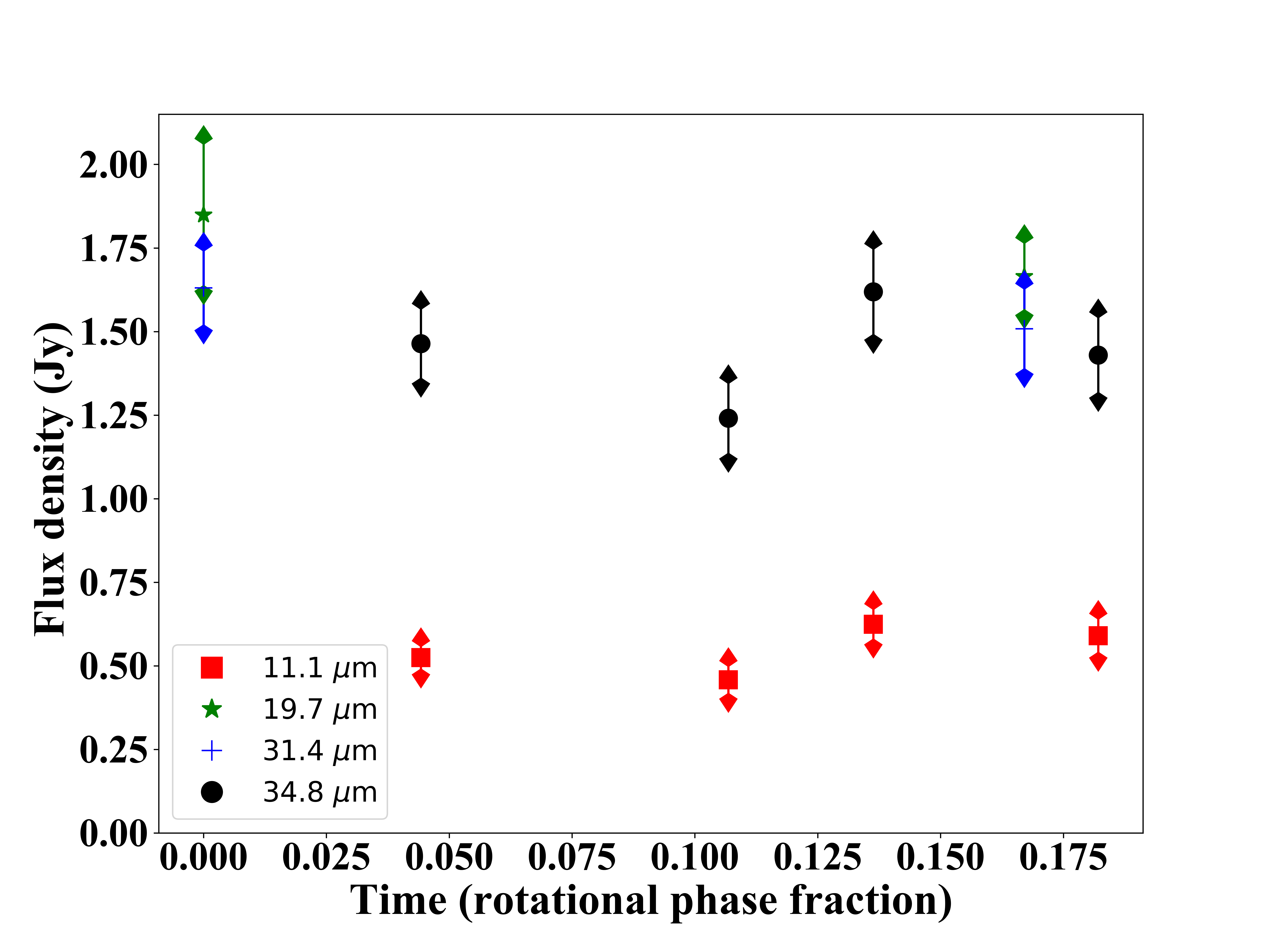}
  \caption{(1162) Larissa mid-infrared photometric data. The graphic shows the flux densities derived from the four wavelength bands of FORCAST, as a function of the rotational phase of the asteroid, together with the given error bars. The obtained mid-infrared photometric light curves follow the trend of increments and decrements along the different wavelengths used in each chop nod shot.}
  \label{larissa_LC}
\end{figure}
\begin{figure}
	\includegraphics[width=\columnwidth]{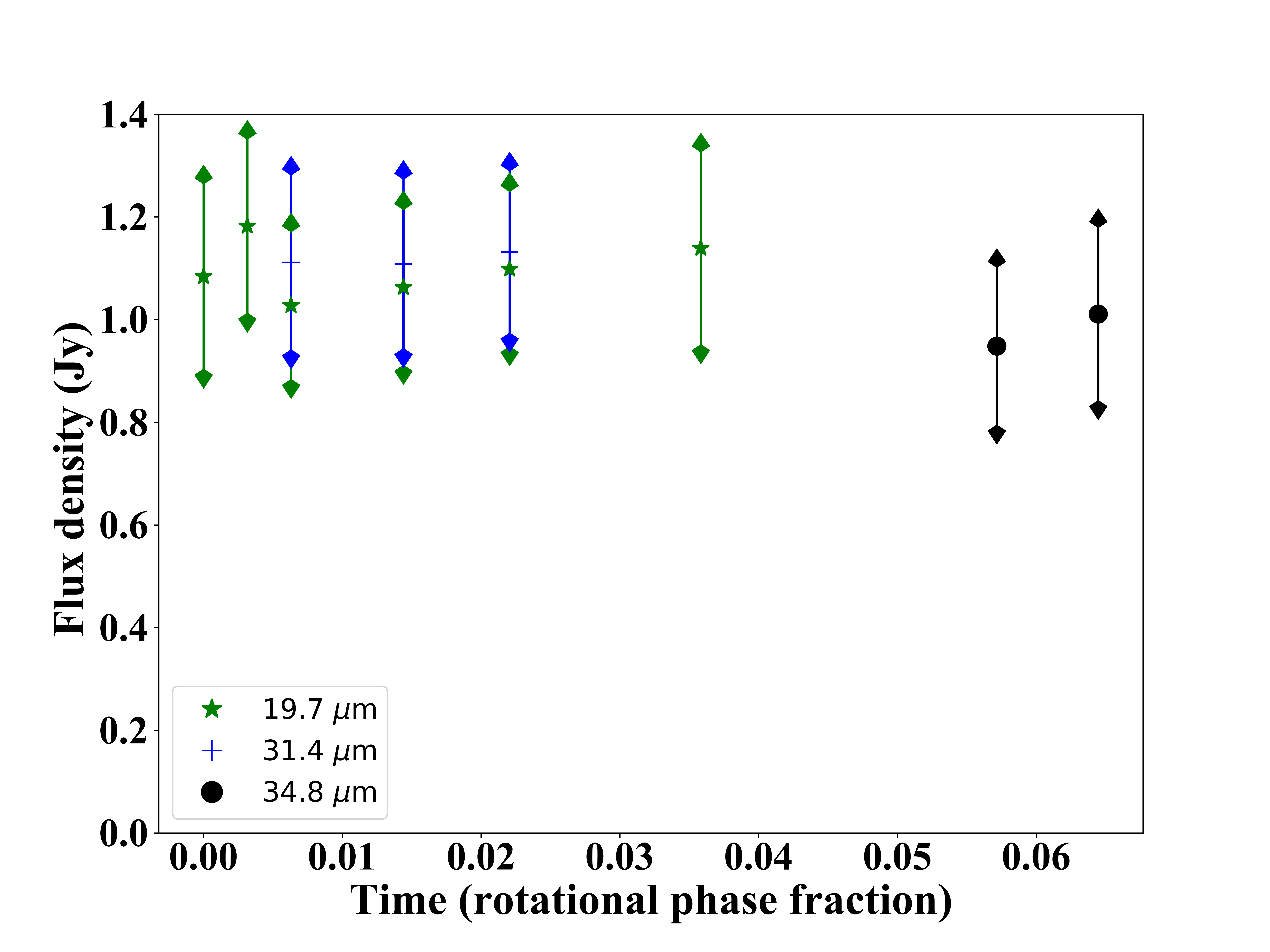}
  \caption{(1911) Schubart mid-infrared photometric data. The graphic shows the flux densities derived from the four wavelength bands of FORCAST, as a function of the rotational phase of the asteroid, together with the given error bars. The obtained mid-infrared photometric light curves follow the trend of increments and decrements along the different wavelengths used in each chop nod shot.}
  \label{schubart_LC}
\end{figure}

We used the Aperture Photometry Tool (APT) \citep{APT} and DS9 \citep{DS9} as complementary tools to decide upon the best values for making the photometry measurements, making particular use of APT functions Aperture Slice, Radial Profile and Curve of Growth, which were combined with the DS9 horizontal and vertical graphs in order to set the correct values for the aperture and annulus extent as required.

We applied appropriate colour correction factors to the measured flux densities, assuming blackbody input spectra for our targets with a temperature close to 150 K, according to the Hildas' distance from the Sun \citep{temperature_asteroids}. The magnitude of these corrections per each waveband filter, however, are values quite close to 1 (between 1.0005 and 1.009), implying a small correction for the raw fluxes calculated, and were taken from Table 3 of \citet{data_red_sofia}. The resulting aperture photometry measurements obtained from the SOFIA images at 11.1, 19.7, 31.5 and 34.8 $\mu$m are presented in Figures \ref{larissa_LC} and \ref{schubart_LC}.

To determine the uncertainties in our results, we followed the FORCAST Photometry Recipe\footnote{SOFIA FORCAST Photometry Recipe at {\url{https://www.sofia.usra.edu/sites/default/files/Other/Documents/FORCAST_Photometry_Recipe.pdf}}, accessed on 2020 April 17.}. The error propagation was done by adding the photometry, the calibration with the standard star and the calibration factor in quadrature. In that process, we used values from Table 5 of \citet{data_red_sofia} to provide the appropiate calibration factor, which was 0.055 for (1162) Larissa, and 0.102 for (1911) Schubart.

\section{Thermophysical modelling}
\label{Sec:Thermophysical}
In order to better understand the physical properties of (1162) Larissa and (1911) Schubart, we here present a detailed thermophysical analysis for both objects, based on the observational data we obtained using SOFIA, coupled with archival data taken by \textit{IRAS} \citep{tedesco1986}, \textit{AKARI} \citep{usui} and \textit{WISE} \citep{mainzer}, as detailed in Section \ref{Sec:Observations}.

\begin{figure*}
\centering
\begin{tabular}{cccccc}
\multicolumn{3}{c}{\includegraphics[width=0.5\textwidth,height=0.25\textheight]{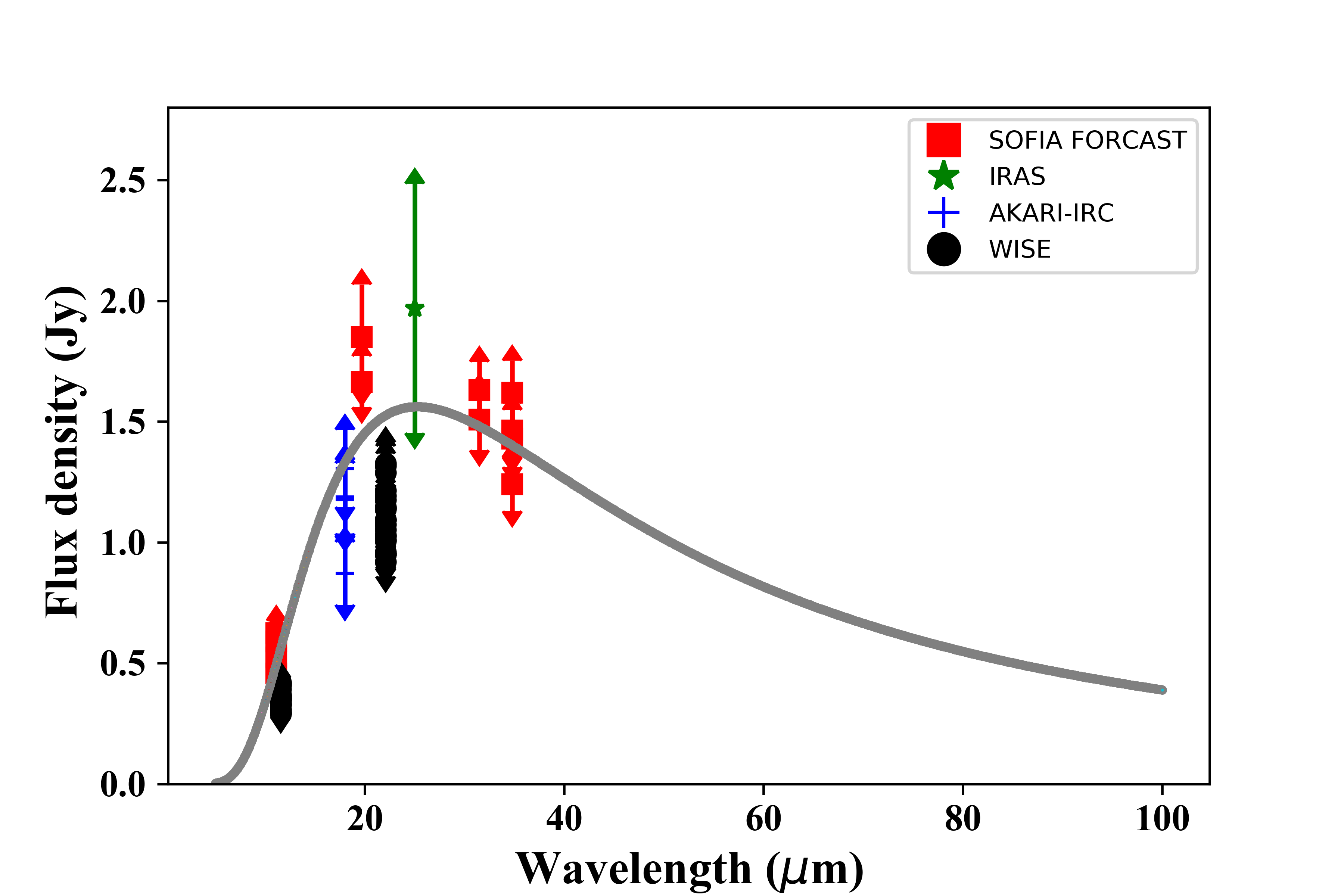}} &\multicolumn{3}{c}{\includegraphics[width=0.5\textwidth,height=0.25\textheight]{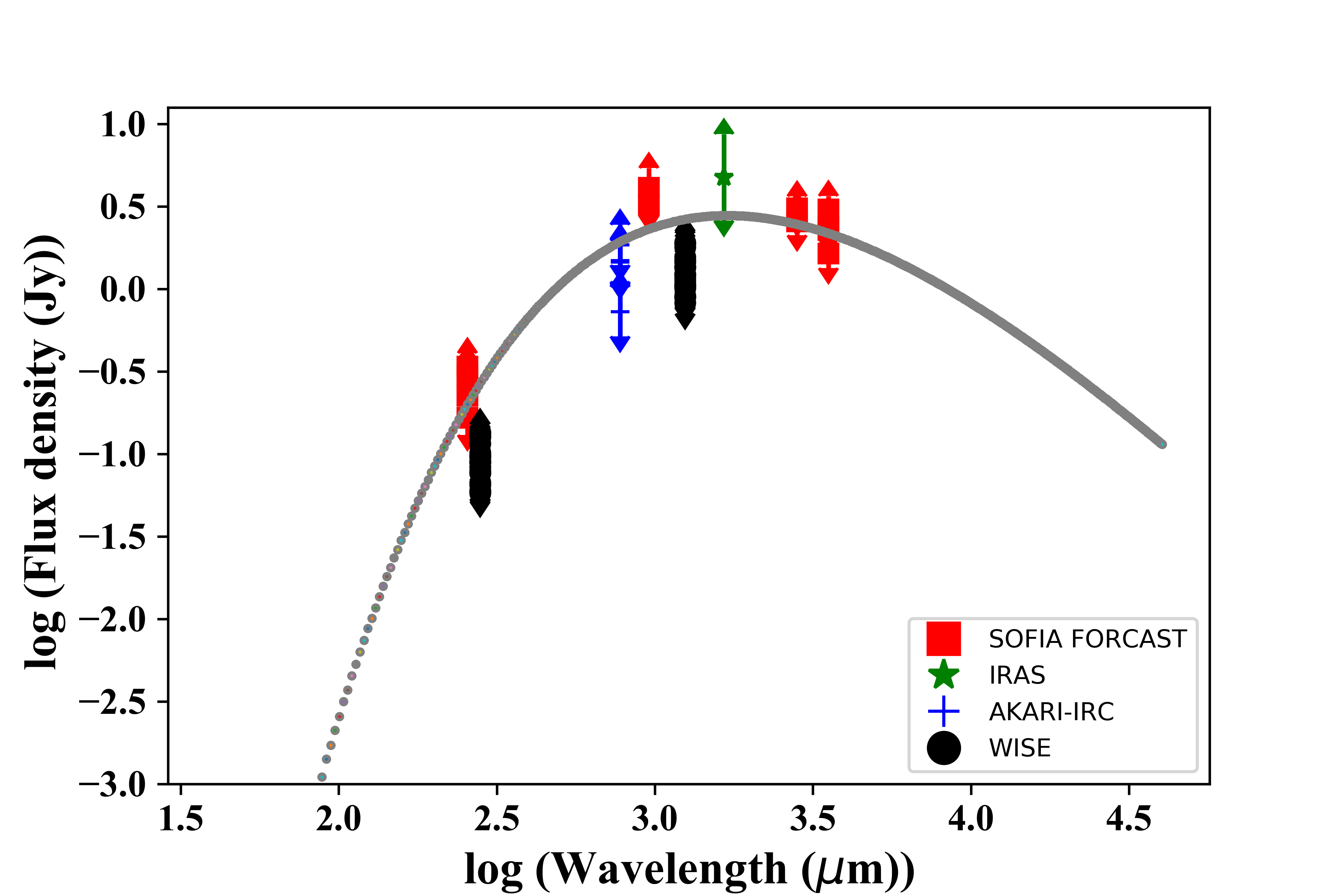}} \\
\multicolumn{3}{c}{\includegraphics[width=0.5\textwidth,height=0.25\textheight]{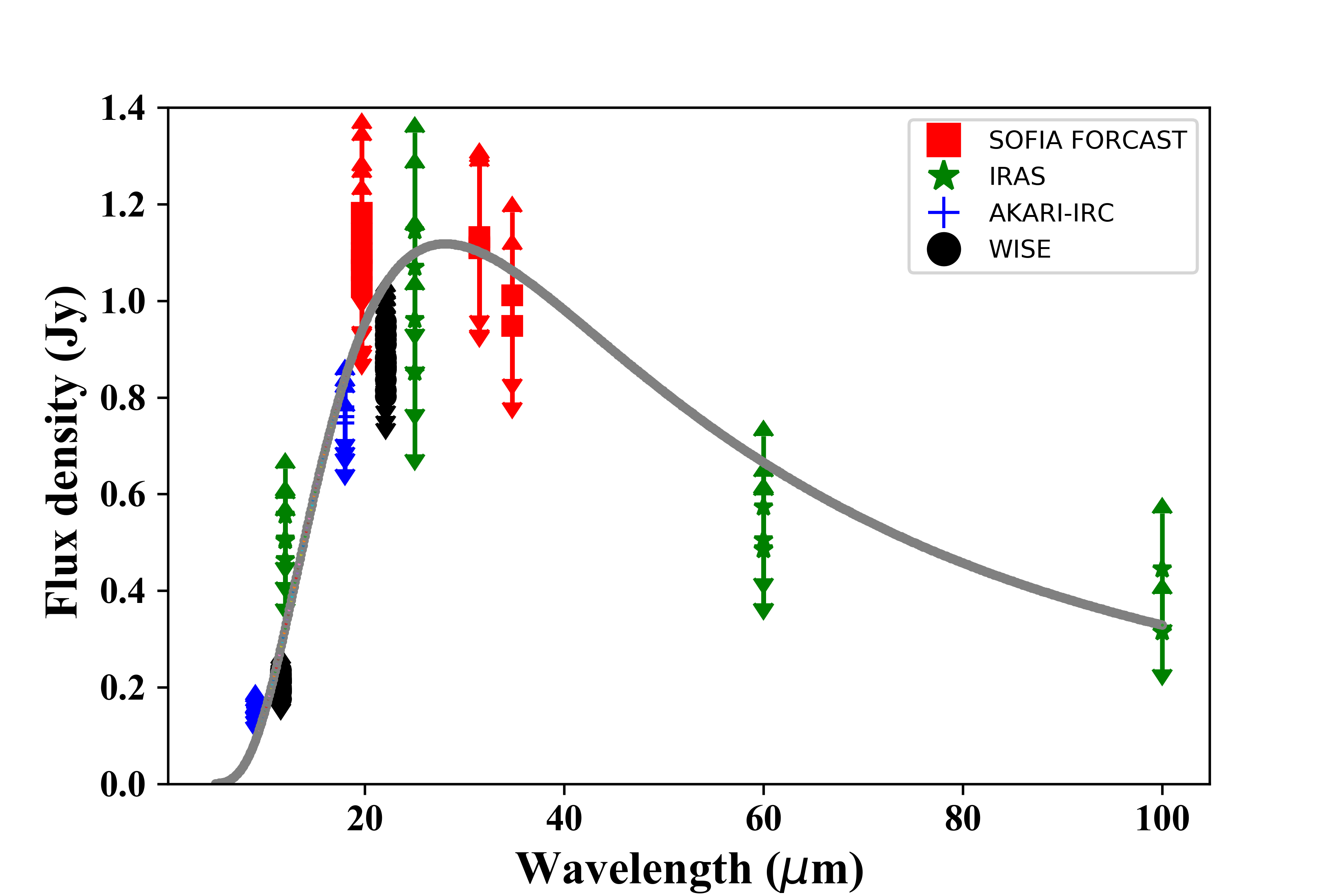}}&\multicolumn{3}{c}{\includegraphics[width=0.5\textwidth,height=0.25\textheight]{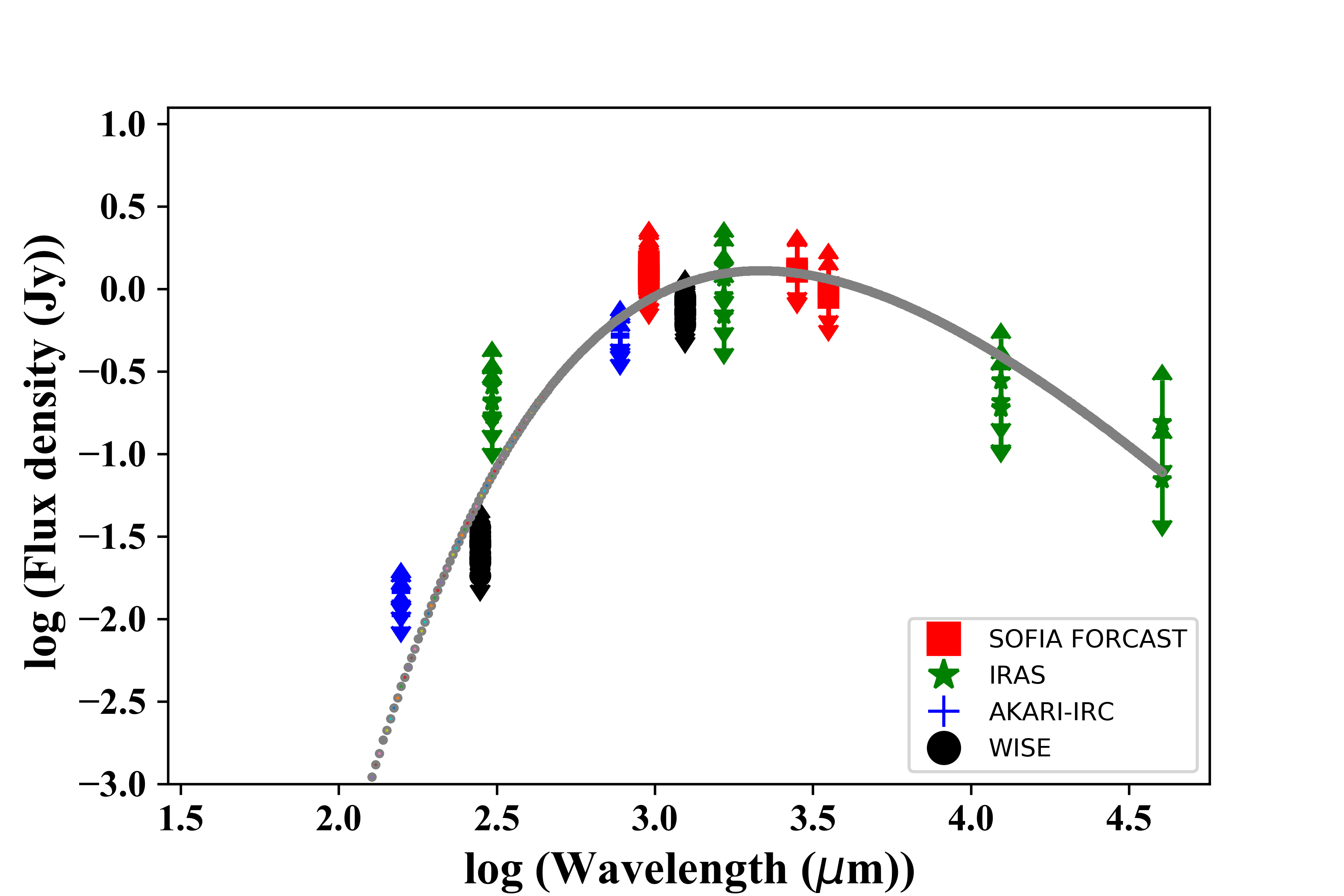}} \\
\end{tabular}
    \caption{Plot showing the both the observations of (1162) Larissa (top) and (1911) Schubart (bottom) and the best fit thermophysical model obtained from those observations. The left plots show the data in linear space, with those on the right showing the same data in a log-log space. The observations are coloured to denote their source - with SOFIA observations shown as red squares, \textit{IRAS} as green stars, \textit{Akari} as blue plus marks, and \textit{WISE} as black circles. For each observation, the vertical lines denote the 1-$\sigma$ uncertainties. The thermophysical model is plotted in grey. \textcolor{black}{The spectral energy distribution was calculated for the SOFIA/FORCAST epoch at lightcurve mid-point, with all the other fluxes scaled to these SOFIA/FORCAST epochs.}}
  \label{obsmodplots}
\end{figure*}

The thermophysical analysis performed in this research is based on the thermophysical model (TPM) detailed in \citet{lagerros,lagerros2,lagerros3} and \citet{Mueller_Lagerros_98, Muller_Lagerros_2002}. In the TPM, the surface temperature is calculated from the energy balance between absorbed Solar radiation, the thermal emission, and heat conduction into the surface material (here, 1-d heat conduction is considered). For a direct comparison with the observed fluxes, we use a given spin-shape solution to calculate the temperature of each surface facet. The integral over all projected surface elements towards the observer then leads to a TPM flux prediction. In principle, we tune the effective size, albedo, and thermal properties, to calculate simultaneously the reflected light (described by the object's absolute magnitude $H$ and the phase slope parameter $G$) and the thermal emission (in direct comparison with the observed thermal emission). This is done for all measurements to find the best-possible solution in size, albedo, thermal inertia, and, if possible, to constrain the object's surface roughness. The results of our analysis are presented in Figure~\ref{obsmodplots}, which shows that, in general, the observational data can be relatively well fit by 
\textcolor{black}{our final thermophysical model solutions, despite the lack of high-quality spin-shape information for both targets.}
\textcolor{black}{For (1162) Larissa, our radiometric tests were performed using spherical and ellipsoidal spin-shape solutions, with a rotation period of 6.519148 hours (\citealt{ivan_larissa,larissa_warner}, J. Durech, private communications). The recent, most likely, pole solutions are (294$^{\circ}$, -28$^{\circ}$) or (114$^{\circ}$, -23$^{\circ}$) in ecliptic longitude and latitude (J. Durech, private communications). In addition (for spherical shapes only), we considered spin-axis orientations of \citep[110$^{\circ}$ ,75$^{\circ}$][]{Cibulkova}\footnote{\textcolor{black}{Using entries from the database at} \url{https://www-n.oca.eu/delbo/astphys/astphys.html}\textcolor{black}{, accessed November 2020.}}, equator-on prograde, equator-on retrograde, pole-on during the SOFIA observations. We assumed that the asteroid had a typical level of surface roughness \citep{Mueller_Lagerros_98}, and solved for its size, albedo and thermal inertia using all SOFIA-FORCAST data (12 points) plus all of the detailed archival data described in section 2.}

\textcolor{black}{Following this methodology, the radiometric range solution for a spherical object, independent of the spin-axis orientation, and assuming H$_{\rm V}$=9.59 and G=0.49 \citep{Oszkiewicz2011} is given by:}
\textcolor{black}{\begin{itemize}
    \item Effective diameter: 41 to 48 km
    \item Geometric V-band albedo 0.10 to 0.15
    \item Thermal inertia below 30 Jm$^{-2}$s$^{-0.5}$K$^{-1}$ 
\end{itemize}}

\textcolor{black}{The reduced {$\chi$}$^{2}$ of the spherical solutions are poor (>4.0). Given an observed (visible) lightcurve amplitude up to 0.22 mag, the poor fit is likely related to our assumption of a simple spherical shape. This is confirmed by deriving radiometric sizes separately for the different data sets. Due to the different viewing geometries for \textit{IRAS}, \textit{Akari}, \textit{WISE}, and SOFIA data, the corresponding dataset-specific sizes deviate substantially and can only be explained by a non-spherical body.}

\textcolor{black}{If we take the two recent spin-pole solutions, we can implement ellipsoidal shape solutions with varying a/b ratios from 1.0 to 1.5 to improve the radiometric analysis. For an a/b ratio of 1.2 we obtained lower {$\chi$}$^{2}$ values (around 2.0 for the (114$^{\circ}$, -23$^{\circ}$) and 2.5 for the (294$^{\circ}$, -28$^{\circ}$) pole solution) with the following values:}
\textcolor{black}{\begin{itemize}
    \item Effective diameter: 46.5$^{+2.3}_{-1.7}$ km
    \item Geometric V-band albedo: 0.12$\pm$~0.02
    \item Thermal inertia: 15$^{+10}_{-8}$ Jm$^{-2}$s$^{-0.5}$K$^{-1}$
    \item Preference for the spin-pole solution with (114$^{\circ}$, -23$^{\circ}$), P\_sidereal = 6.519148 h, ellipsoidal shape with a/b=1.2
\end{itemize}
}

\textcolor{black}{These two spin-poles with the specified ellipsoidal shape have the big advantage that they can reproduce reasonably well the observed thermal lightcurve (22 data points taking in Jan 2010, and another 20 data points taken in July 2010, each time within about one day), and the overall flux changes with wavelengths and phase angles. But based on the thermal data alone, it is not possible to entirely exclude other spin-shape solutions. However, future observations of the asteroid will prove critical in order to mitigate the remaining spin-shape uncertainties and to improve the radiometric solution further.}

\textcolor{black}{In the case of (1911) Schubart, using the multiple thermal data described in section~\ref{Sec:Observations}, in combination with a spherical shape, a rotation period of 11.9213 h \citep{schubart_2016_ok}, and pole solution of (320$^{\circ}$, 20$^{\circ}$) (J. Durech, private communication) leads to acceptable {$\chi$}$^{2}$ values below 2.0 for all spin-axis orientations. The data are reasonably well fitted, only the \textit{IRAS} 100 $\mu$m observations seem to be problematic. The derived radiometric size range is between about 64 km up to 78 km, the corresponding albedos are between 0.033 and 0.42, respectively. However, the analysis process gave a slight preference for a spin-axis with high-obliquity like the pole-on during SOFIA observations or one of the recently estimated solutions from lightcurve inversion techniques.}

For completeness, we performed additional tests considering the alternative rotation period of 7.9121 hours rotation period proposed in \citet{schubart_2017}. Changing the rotation period in this manner had no impact on the results, showing no influence on the spin-size-albedo-TI solution. However, we find that the 11.9 h rotation period seems to explain the \textit{WISE} data in a more consistent way.

\textcolor{black}{We also investigated if ellipsoidal shape solutions with varying a/b ratios (from 1.0 to 1.5) lead to better fits of all data. An improvement is seen only in the high-obliquity case (with the asteroid seen pole-on during SOFIA observations). Here, an axis ratio a/b=1.3 pushes the {$\chi$}$^{2}$ close to 1.0. However, the \textit{WISE} lightcurve data set (36 data points taken within 3.6 days in April 2010) is still not very well matched. This points to a more complex shape solution. But if we accept this best-fit pole-on spin-shape solution as baseline, then our radiometric solution provides the following values:}
\textcolor{black}{\begin{itemize}
    \item Effective diameter: 72$^{+3}_{-4}$ km
    \item Geometric V-band albedo: 0.039$\pm$~0.02
    \item Thermal inertia: 10$^{+20}_{-5}$ Jm$^{-2}$s$^{-0.5}$K$^{-1}$
\end{itemize}}

\begin{figure*}
\begin{tabular}{cc}
    \includegraphics[width=\columnwidth]{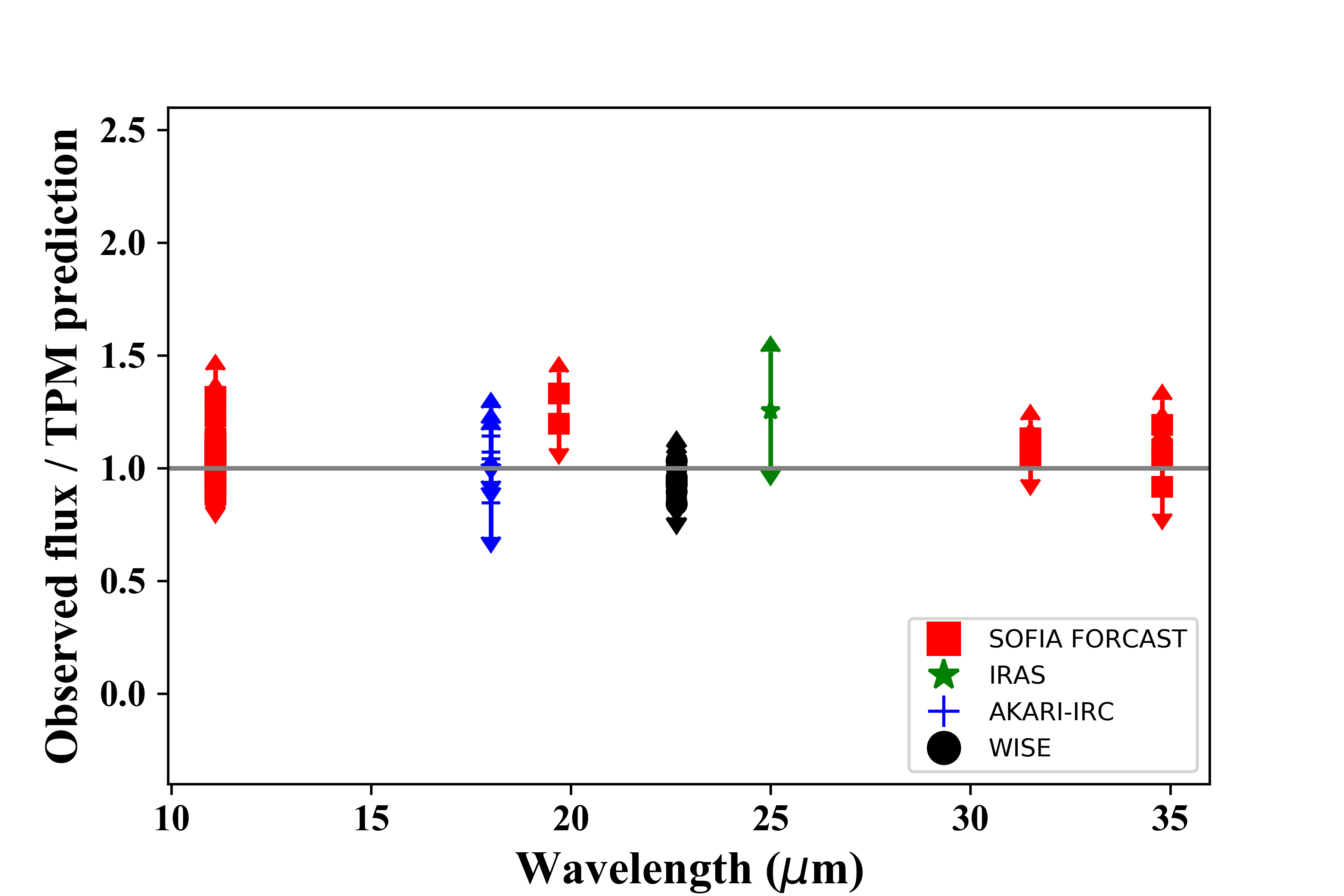}
    \includegraphics[width=\columnwidth]{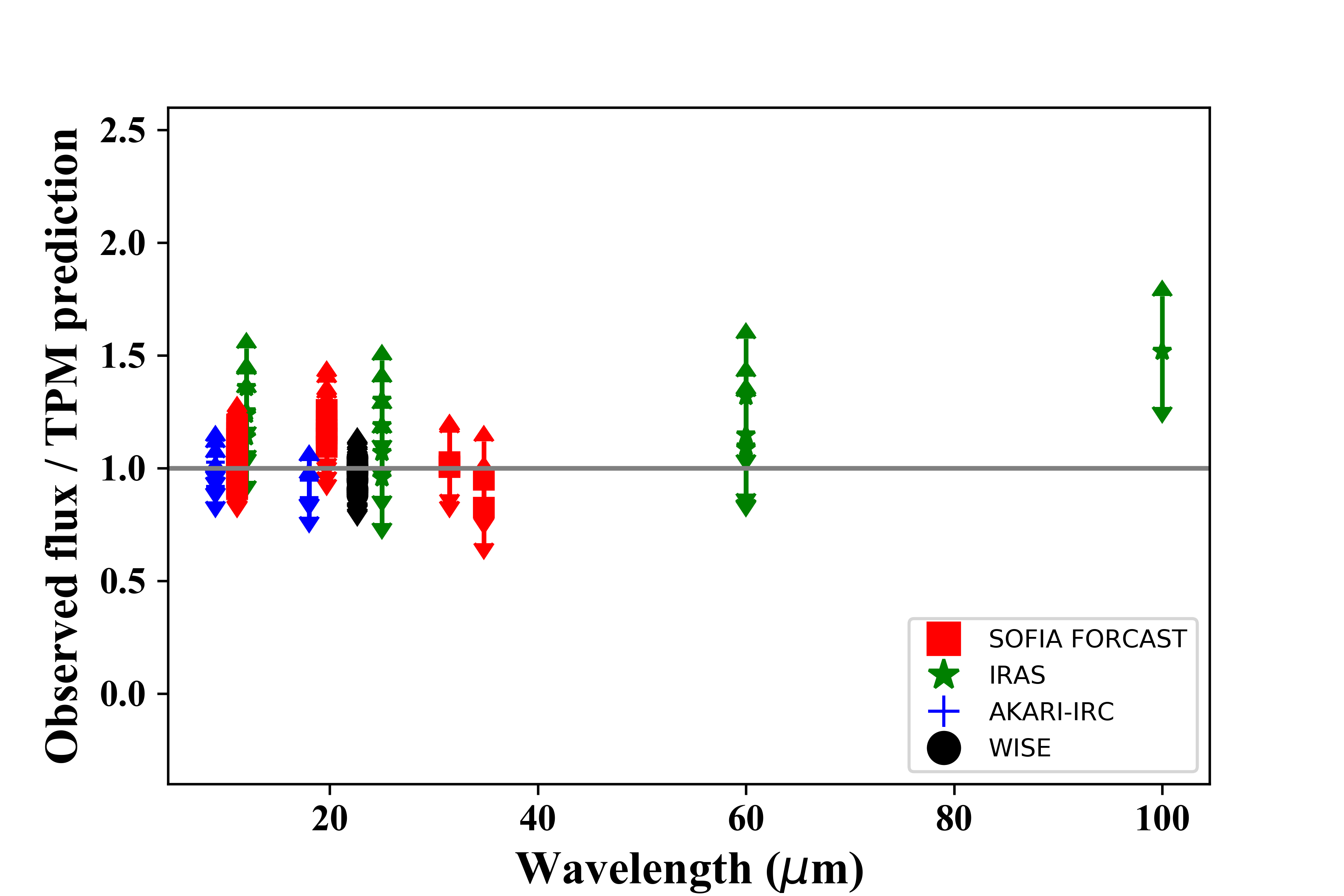}\\
    \includegraphics[width=\columnwidth]{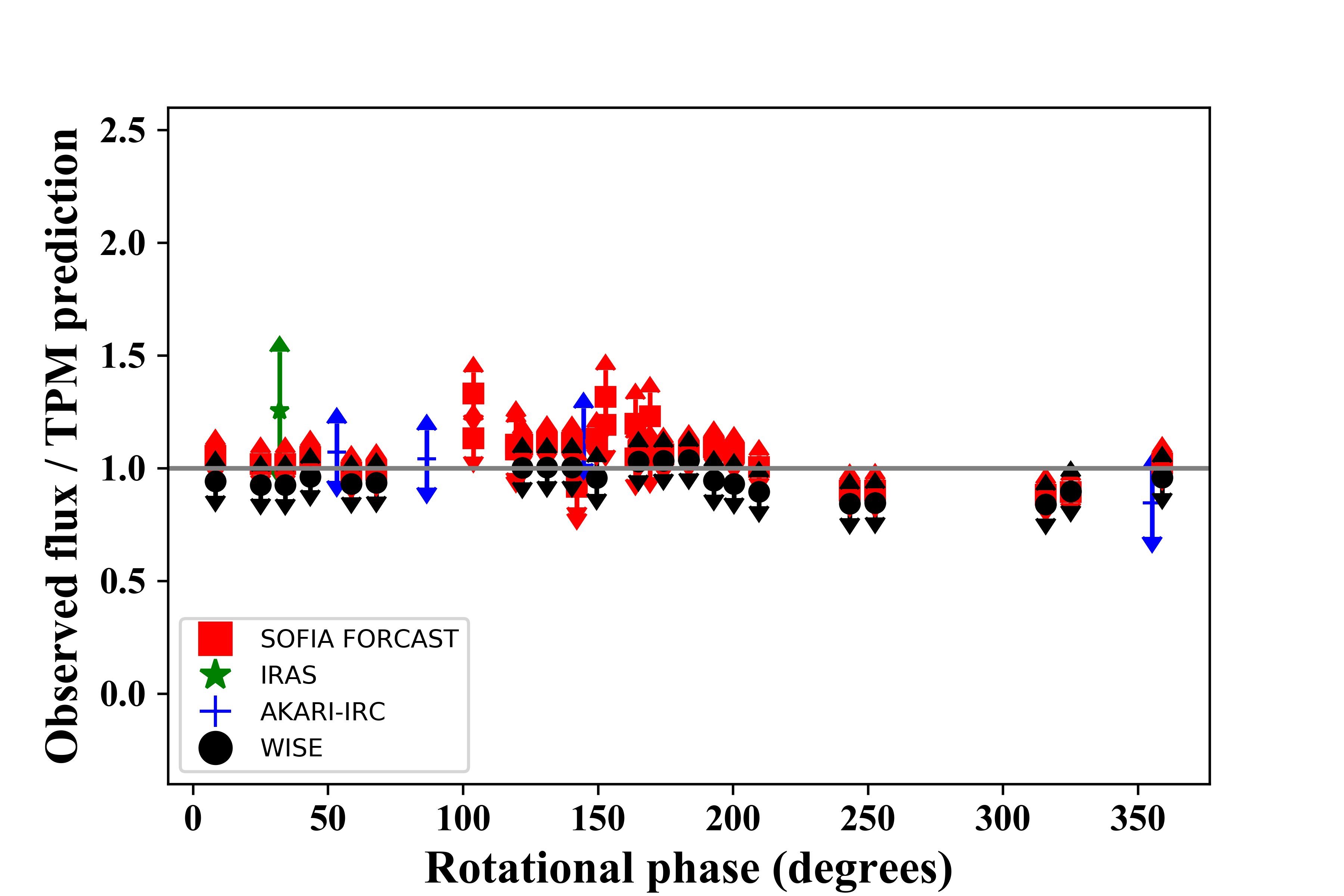}
    \includegraphics[width=\columnwidth]{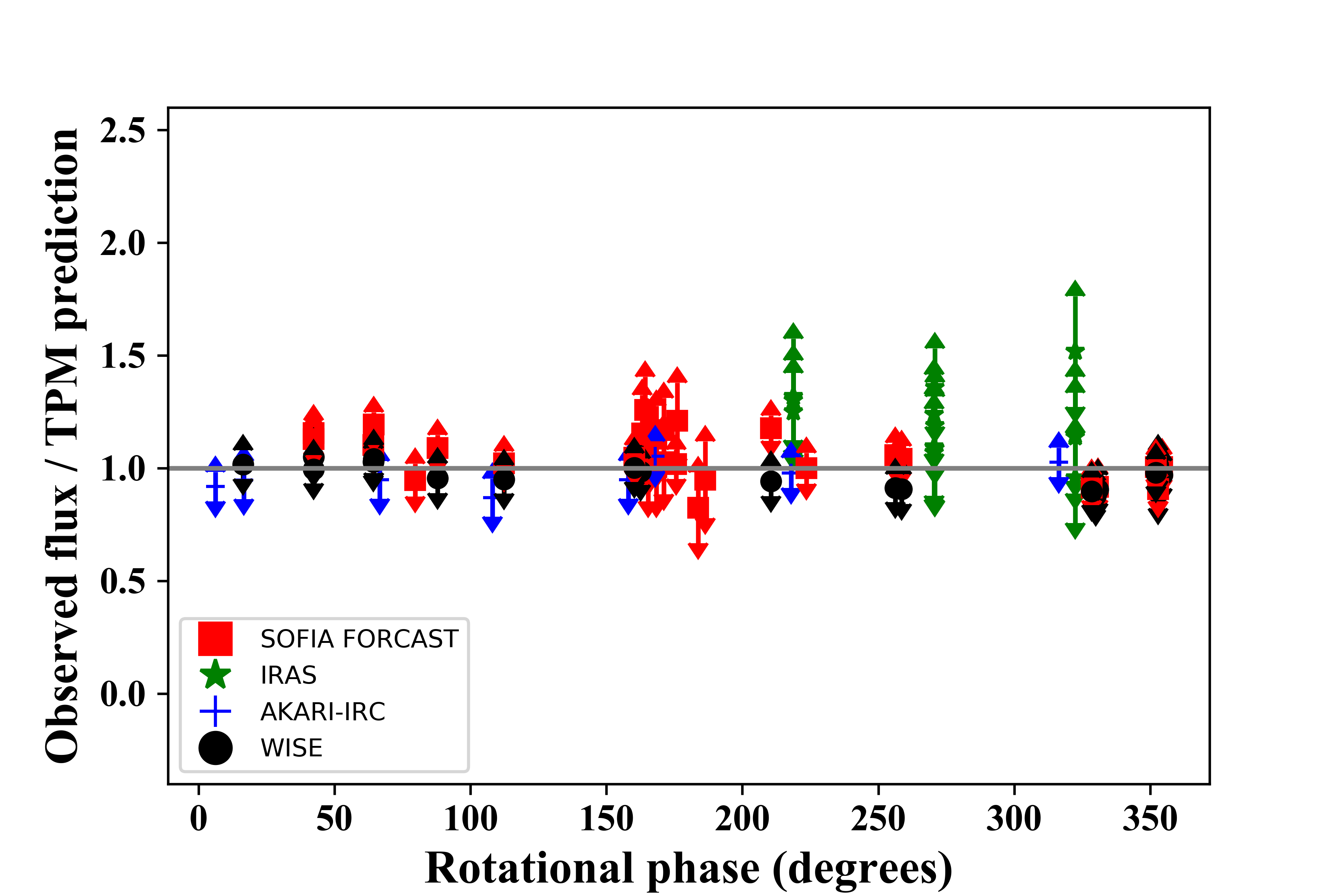}\\
\end{tabular}
  \caption[width=\columnwidth]{The degree to which the thermophysical models agree with the observations can be demonstrated by taking the observed flux and dividing it by the predicted values. Here, we show the degree to which the observations and model agree for (1162) Larissa (left) and (1911) Schubart (right). The upper panels show the agreement between the model and observations as a function of wavelength, whilst the lower panel plots the agreement instead as a function of rotational phase for the two asteroids. As in Figure~\ref{obsmodplots}, the observations are coloured to denote which instrument obtained them. In general, the observations and model are in good agreement, albeit with noticeable scatter at any given wavelength.}
  \label{OBSMOD_residual}
\end{figure*}

\textcolor{black}{In Figure~\ref{OBSMOD_residual}, we show (in the upper panels) the degree to which the thermophysical model agrees with our observations, as a function of wavelength. Aside from the 100 $\mu$m observations of (1911) Schubart made by \textit{IRAS}, it is clear that, in general, the model and data are in good agreement. The observation-to-model ratios are well balanced over a wide range of wavelengths, phase angles, and rotational phases.}

\textcolor{black}{In the lower panels of Figure~\ref{OBSMOD_residual}, we plot the agreement between our thermophysical model and the gathered observational data against rotational phase. For both targets, there appears to be a remaining small variation in goodness of fit between our model and the data as a function of rotational phase. In general, there is still a significant scatter in the observational-to-model ratios, which lends further support to the idea that the two asteroids may well have more complex shapes.}

\begin{center}
\begin{table*}
 \begin{tabular}{ccccc}
\toprule
 &\multicolumn{2}{c}{(1162) Larissa}
&
\multicolumn{2}{c}{(1911) Schubart} \\\cmidrule(r){2-3}\cmidrule(l){4-5}
Variables &\textbf{Published data}&\textbf{Current research}  &\textbf{Published data}&\textbf{Current research}   \\ 
\hline
Effective diameter [km] &40.38-52.16 &$46.5^{+2.3}_{-1.7}$ &64.66-92.37 &72$^{+3}_{-4}$\\ \\
Geometric albedo in V-band &0.102-0.186 &0.12~$\pm$~0.02 &0.022-0.04 &0.039~$\pm$~0.05\\ \\
Thermal inertia [Jm$^{-2}$s$^{-0.5}$K$^{-1}$] &No value &$15^{+10}_{-8}$ &No value &$10^{+20}_{-5}$\\
\hline
\bottomrule
\end{tabular}
\caption{Results of thermophysical modelling for (1162) Larissa and for (1911) Schubart. This Table collates data from \citet{Tedesco2004}, \citet{usui}, \citet{Grav2012}, \citet{Nug15}, \citep{alilagoa} and \citet{primass}}
\label{results_table}
\end{table*}
\end{center}

\section{Dynamical modelling}
\label{Sec:Dynamics}

To complement our study of the physical properties of (1162) Larissa and (1911) Schubart, we performed two detailed suites of $n$-body simulations to examine the long-term dynamical evolution of the two asteroids. For each asteroid, we used the Hybrid integrator within the $n$-body dynamics package \textsc{Mercury} \citep{mercury} to follow the evolution of a swarm of 61\,875 test particles for a period of $10^9$ years, under the gravitational influence of the Sun, Jupiter, Saturn, Uranus and Neptune. An integration time step of 60 days was used, and each individual particle was followed for the full $10^9$ year duration of the simulations, unless it collided with one of the four giant planets, the Sun, or was ejected to a heliocentric distance of 1000 au.

The test particles themselves were generated in a manner similar to that used in our earlier work \citep[e.g.][]{QR322,LC18,Anchises}, with ``clones'' generated in a regular 4-dimensional grid spanning the full $\pm$3$\sigma$ uncertainties in the objects semi-major axis, $a$, eccentricity, $e$, inclination, $i$, and mean anomaly, $M$. In total, 25 unique values of semi-major axis were tested. For each of these values, 15 unique values of eccentricity were examined, creating a $25 \times 15$ grid in $a-e$ space. For each $a-e$ ordinate, 15 unique inclination values were tested - with 11 unique mean anomalies being considered for each individual $a-e-i$ ordinate. In this way, we generated a four-dimensional hypercube of test particles, distributed on an even lattice, with dimensions $(25 \times 15 \times 15 \times 11)$ in $(a,e,i,M)$ space. The initial orbital elements for the two Hildas used as the basis for our integrations, along with their associated uncertainties, are detailed in Table~\ref{elements}.

\begin{figure*}
\hspace*{-0.5cm}\begin{tabular}{cccc}
 \includegraphics[width=0.24\textwidth,height=0.24\textwidth]{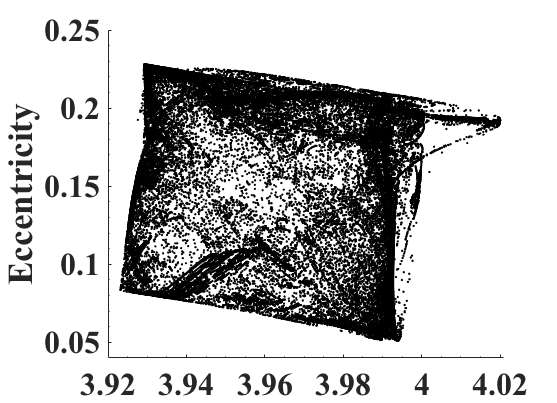} & \includegraphics[width=0.24\textwidth,height=0.24\textwidth]{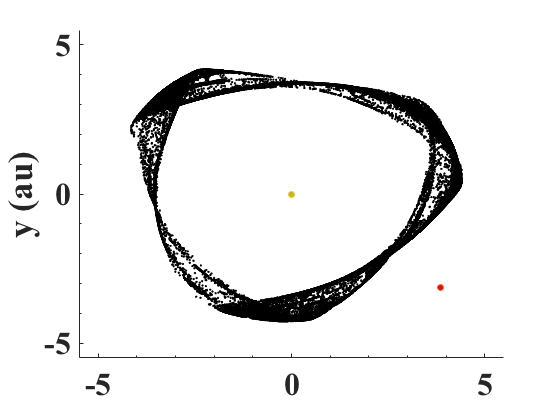} & \includegraphics[width=0.24\textwidth,height=0.24\textwidth]{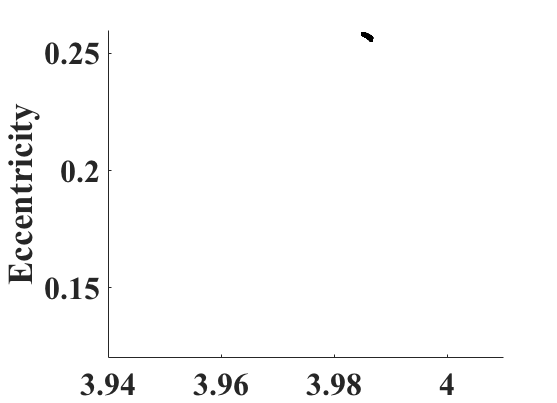} & \includegraphics[width=0.24\textwidth,height=0.24\textwidth]{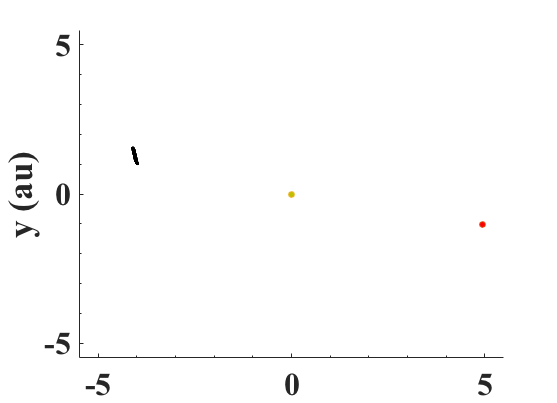}\\
  \includegraphics[width=0.24\textwidth,height=0.24\textwidth]{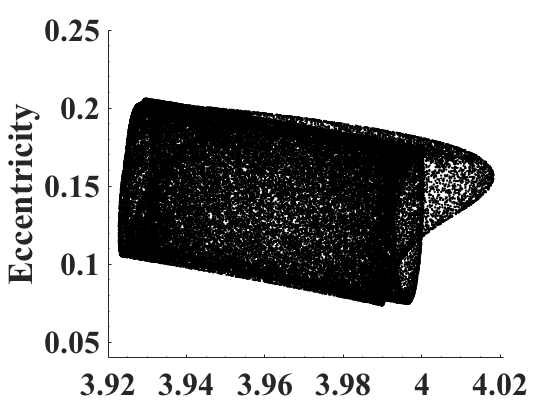} & \includegraphics[width=0.24\textwidth,height=0.24\textwidth]{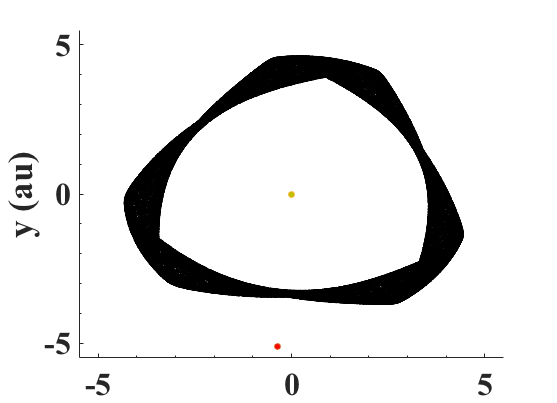} & \includegraphics[width=0.24\textwidth,height=0.24\textwidth]{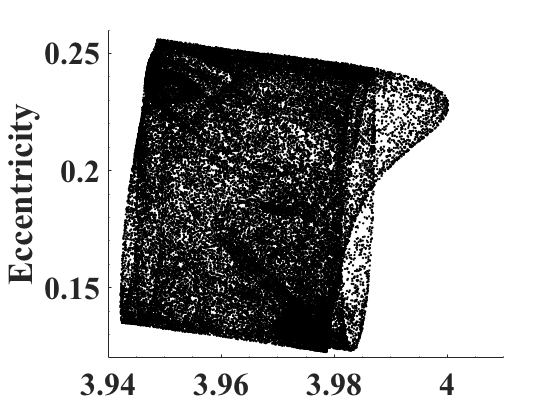} & \includegraphics[width=0.24\textwidth,height=0.24\textwidth]{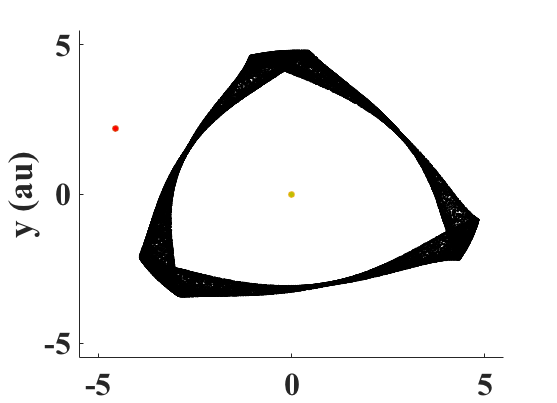} \\
 \includegraphics[width=0.24\textwidth,height=0.24\textwidth]{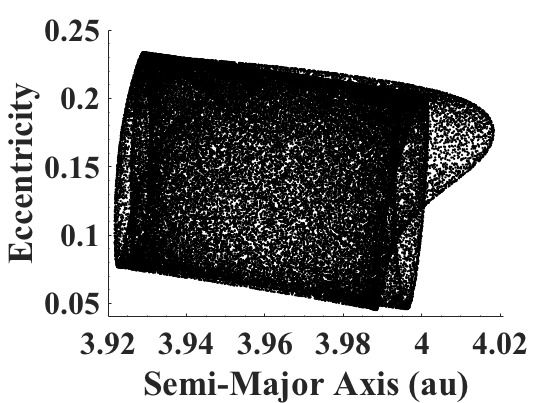} & \includegraphics[width=0.24\textwidth,height=0.24\textwidth]{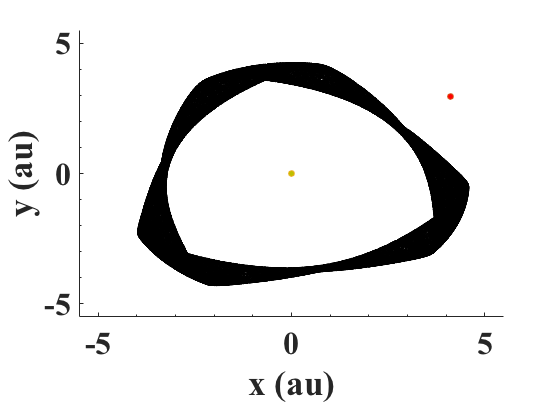} & \includegraphics[width=0.24\textwidth,height=0.24\textwidth]{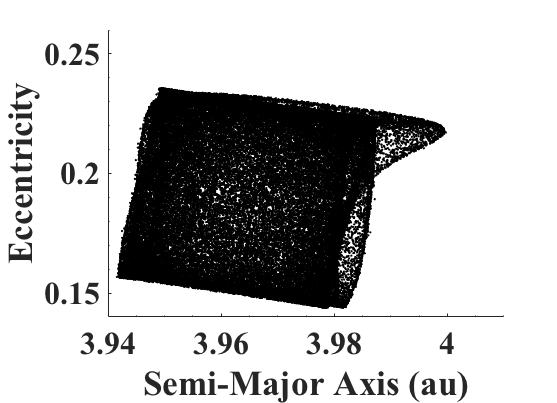} & \includegraphics[width=0.24\textwidth,height=0.24\textwidth]{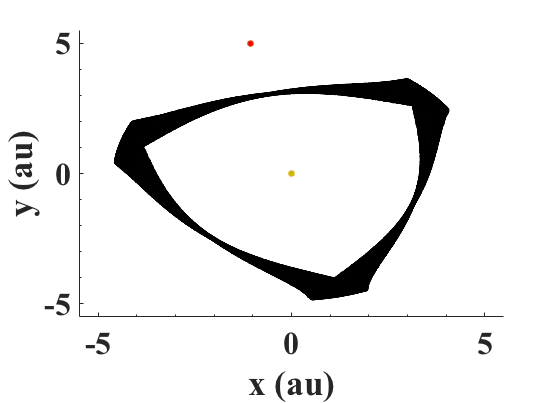} \\
\end{tabular}
\caption{The dispersion of a suite of 61\,875 clones of (1162) Larissa (left two columns) and (1911) Schubart (right two columns). The top row shows the clones after 10 Myr of integration time have elapsed, the central row shows them after 100 Myr, and the lower row shows them after 1 Gyr, at the end of our simulations. For each object, the left-hand column shows the clones in semi-major axis-eccentricity space, whilst the right shows them in plan view, looking down at the Solar system from above. In the plan view plots, the yellow point at the centre denotes the location of the Sun, whilst the red dot shows the location of Jupiter at the instant the clones are plotted. Whilst the clones of both (1162) Larissa and (1911) Schubart take some time to disperse, it is noticeable that the clones of (1911) Schubart do so more slowly, and over a smaller distance in orbital element space.}
\label{hildatrig}
\end{figure*}

In stark contrast to our previous studies of resonant Solar system small bodies \citep[][]{QR322,LC18,Anchises}, both (1162) Larissa and (1911) Schubart proved to exhibit staggering dynamical stability. Over the course of our simulations, not a single clone of either object escaped from the Hilda population. Over time, the test particles did diffuse to fill the entirety of the ``Hildan Triangle'' -- the striking feature in plan view of the Solar system occupied by the Hildas (as shown in Figure~\ref{hildatrig}) -- but none became sufficiently excited to escape that population.

The extreme stability of the two Hildas argues towards them being primordial, rather than captured objects. It stands as an interesting contrast to the dynamical evolution of objects studied in a similar manner in the Jovian and Neptunian Trojan populations - all of which exhibit some degree of instability \citep[e.g.][]{LevTroj,MS02,QR322,LC18,Anchises,DiSisto19,Holt20}. Given the extreme stability of these two objects, it seems reasonable to suggest that future studies of the migration of the giant planets and their impact on the Solar system's young small body populations should examine whether those models can replicate such tightly captured objects moving on dynamically cold orbits in the Hilda region.

\begin{center}
\begin{table*}
\begin{tabular}{ccccc}
\toprule
&\multicolumn{2}{c}{1162 Larissa}
&
\multicolumn{2}{c}{1911 Schubart} \\\cmidrule(r){2-3}\cmidrule(l){4-5}
Variables &\textbf{Best fit value}&\textbf{1-$\sigma$ uncertainty}  &\textbf{Best fit value}&\textbf{1-$\sigma$ uncertainty}   \\
\hline
Semi-major axis [au]  & 3.9234  & 5.092 $\times~10^{-8}$  & 3.97885 & 4.723 $\times~10^{-8}$ \\
Eccentricity      & 0.110524 & 6.291 $\times~10^{-8}$  & 0.165839 & 6.74 $\times~10^{-8}$ \\
Inclination [$^\circ$]  & 1.888   & 6.972 $\times~10^{-6}$  & 1.651  & 7.72 $\times~10^{-6}$ \\
$\omega$ [$^\circ$]   & 39.791  & 1.825 $\times~10^{-4}$   & 285.447 & 2.269 $\times~10^{-4}$ \\
$\Omega$ [$^\circ$]   & 212.222  & 1.847 $\times~10^{-4}$   & 183.727 & 2.282 $\times~10^{-4}$ \\
Mean Anomaly [$^\circ$] & 59.024  & 3.447 $\times~10^{-5}$  & 246.572 & 2.093 $\times~10^{-5}$ \\
\hline
\bottomrule
\end{tabular}
\caption{Orbital elements, and their associated uncertainties, for (1162) Larissa and (1911) Schubart, as taken from the AstDys database on January 12, 2012.}
\label{elements}
\end{table*}
\end{center}

\section{Discussion and conclusions}
\label{Sec:Conclusions}

In this work, we have presented a detailed thermophysical and dynamical analysis of the Hilda asteroids (1162) Larissa and (1911) Schubart, using observations made using the FORCAST instrument on the SOFIA airborne observatory on 2011 June 4 and June 8, supplemented by archival data from the \textit{IRAS}, \textit{AKARI} and \textit{WISE} space observatories. Our dynamical simulations reveal both Hildas to exhibit extreme dynamical stability, with not a single one of the 61\,875 clones generated of each object escaping from the Hilda region in the $10^9$ years of our integrations. This finding is strongly suggestive that the two objects are primordial in nature, rather than having been captured, since objects in populations that are widely accepted to have been captured during the Solar system's youth tend to exhibit dynamical instability to some degree \citep[e.g.][]{LevTroj,MS02,QR322,LC18,Anchises,DiSisto19,Holt20}.

Our thermophysical analysis of the two Hildas reveals that both asteroids have non-spherical shapes, with the physical properties we derive on the basis of our analysis being in broad agreement with previously published values, albeit with markedly reduced uncertainties (See Table \ref{results_table}).

For (1162) Larissa, the geometric V-band albedo we obtained (0.12~$\pm$~0.02) is in close agreement with the 0.127 $\pm$ 0.009 given by \citet{usui}. It does not, however, agree with the values of 0.169 $\pm$ 0.012 and 0.186 $\pm$ 0.03 given by \citet{primass} and \citet{Grav2012}.
Despite this, our results confirm that (1162) Larissa is amongst the most reflective Hildas, whose mean albedo (0.055) is markedly darker \citep{primass}.
In this sense, it sounds reasonable to have slightly different values regarding surveys that use another infrared wavelengths to study asteroids and thus may under or over estimate the geometric albedo, making valuable the effort for making use of instrumentation such as FORCAST operating at 3 to 50 $\mu$m, which spans the wavelength range of peak thermal emission from asteroids \citep{asteroids_thermal_IR,thermal_peak}.
Our results yield a refined value for the effective diameter of (1162) Larissa of $46.5^{+2.3}_{-1.7}$ km, a measurement that sits close to the centre of the range of previously published diameters, spanning 40.38 to 48.59 km \citep{Tedesco2004,usui,Nug15}. 

\textcolor{black}{Our thermophysical analysis of (1911) Schubart proved more challenging, as a result of having to discard a number of our images of the asteroid}. When our observations are taken in concert with those obtained by \textit{WISE}, we find better agreement with the longer rotation period of 11.915 hours proposed in \citep{schubart_2017}, rather than the short 7.9 hour period also presented in that work. By incorporating that longer period into our analysis, we obtain a geometric V-band albedo for (1911) Schubart of 0.039$\pm$ 0.05, which is in agreement with the mean value of 0.035 presented by \citet{Grav2012} and \citet{Nug15}. This confirms that (1911) Schubart is an unusually dark asteroid, amongst the least reflective objects known in the Solar system. Indeed, this outcome is in strong agreement with the work of \citet{hidas-romanishin}, who state that (1911) Schubart dynamical family as a whole is 30 per cent darker than the bulk of the Hilda population outside that family. Our results yield an effective diameter for (1911) Schubart of $72^{+3}_{-4}$ km. This result is once again in broad agreement with those published in the literature, which range between 64.66 and 92.37 km \citep{Tedesco2004,usui,Nug15}.

Our thermophysical analysis also enabled us to determine the thermal inertias of the two Hilda asteroids we studied. That analysis yielded values of \textcolor{black}{15$^{+10}_{-8}$ Jm$^{-2}$s$^{-0.5}$K$^{-1}$ for (1162) Larissa and 10$^{+20}_{-5}$ Jm$^{-2}$s$^{-0.5}$K$^{-1}$ for (1911) Schubart}. These results lie within the broad range of values given by \citet{Mueller_Lagerros_98} in their study of a number of the largest main-belt asteroids, whose thermal inertias were found to range between 5 and 25 Jm$^{-2}$s$^{-0.5}$K$^{-1}$, and stand in stark contrast to the outliers found in previous studies of inner main belt asteroids, such as (306) Unitas, which yielded values as high as 260 Jm$^{-2}$s$^{-0.5}$K$^{-1}$, and (277) Elvira, with values reaching 400 Jm$^{-2}$s$^{-0.5}$K$^{-1}$ \citep{Delbo2015}. 

Flux variations in the thermal measurements provide some evidence that (1911) Schubart has an elongated shape since the spherical 
\textcolor{black}{one} produces significant variations in the observation-divided-by-model plot which seem to correlate with the most-likely rotation period. Future improved spin-shape solution might resolve this issue and will allow to refine the radiometric solutions for both Hildas.

It is interesting that the two asteroids studied, (1162) Larissa and (1911) Schubart, are similar in size and thermal inertia, but so different in terms of albedo. Indeed, they seem to mark extremes in the observed properties of the Hildas - with (1162) Larissa being unusually reflective, and (1911) Schubart particularly dark.

These albedo values motivate us to review the taxonomy of both asteroids; the P-type assigned to (1162) Larissa in the (LCDB){\footnote{Asteroid Light Curve Data Base at {\url{http://www.minorplanet.info/PHP/lcdbsummaryquery.php}}, accessed on 20th March 2020.}} and JPL{\footnote{NASA/JPL Small-Body Database at {\url{https://ssd.jpl.nasa.gov/sbdb.cgi}}, accessed on 20th March 2020.}} is not in tune with its low thermal inertia of \textcolor{black}{15$^{+10}_{-8}$ Jm$^{-2}$s$^{-0.5}$K$^{-1}$}, suggesting a clear difference in physical nature compared to the P-type low albedo (1911) Schubart and its \textcolor{black}{10$^{+20}_{-5}$ Jm$^{-2}$s$^{-0.5}$K$^{-1}$}.

\citet{hidas-romanishin} suggest that the (1911) Schubart family are interlopers to the Hilda region. The low albedo we obtain from our observations supports the idea that (1911) Schubart is physically somewhat different to the other Hildas -- but the question remains of whether those differences are the result of the collision that formed the (1911) Schubart family, or are instead indicative that the parent of the family is an interloper in the Hilda region. Our dynamical results reveal that (1911) Schubart is extremely firmly embedded in the Hilda population, trapped securely in 3:2 mean-motion resonance with Jupiter -- a result that seems hard to reconcile with its having been captured to the population after the bulk had already formed. This result is potentially supported by the similarity between the thermal inertias of the two objects -- suggesting that the divergent albedos are the result of processes that have occurred through the lifetime of the objects, rather than being indicative of their having markedly different origins. 

It is clear that further study is needed before we can conclusively determine whether any or all of the Hildas are captured objects, rather than having formed in-situ. The origin of the Hildas is still an open and crucial question, and it seems likely that the answer to that question will shed new light on the formation and evolution of the Solar system as a whole.

\section*{Acknowledgements}

This research is mainly based on observations taken by the NASA/DLR SOFIA airborne telescope. SOFIA is jointly operated by the Universities Space Research Association, Inc. (USRA, NASA contract NAS2-97001), and the Deutsches SOFIA Institut (DSI) under DLR contract 50 OK 0901, to the University of Stuttgart. 
This work has made use of the NASA/JPL Solar System Dynamics page, at \url{https://ssd.jpl.nasa.gov/}. It has also made use of the AstDyS database, at \url{https://newton.spacedys.com/astdys/index.php?pc=3.0}, and NASA’s Astrophysics Data System, ADS.
Also, this research made use of Astropy, a community-developed core Python package for Astronomy, Photutils, an affiliated package of Astropy for detecting and performing photometry of astronomical sources, and Aperture Photometry Tool (APT), a software created to perform simple and effective aperture photometry analyses.

TM has received funding from the European Union’s Horizon 2020 Research and Innovation Programme, under Grant Agreement No. 687378, as part of the project "Small Bodies Near and Far" (SBNAF). JPM acknowledges research support by the Ministry of Science and Technology of Taiwan under grants MOST104-2628-M-001-004-MY3, MOST107-2119-M-001-031-MY3, and MOST109-2112-M-001-036-MY3, and Academia Sinica under grant AS-IA-106-M03. 

\textcolor{black}{CCh wishes to acknowledge the fruitful discussions in the field of Infrared Astronomy with Dr. Felipe Barrientos.}

The authors wish to thank the anonymous referee, whose comments helped to improve the flow and clarity of this work.

\textit{Facilities:} SOFIA (FORCAST)

\textit{Software:} Astropy \citep{2013AstroPy,2018AstroPy}, Numpy \citep{2006Oliphant,2011numpy,2020numpy}, Matplotlib \citep{2007Hunter}, \textsc{Mercury} \citep{mercury}.\\

\section*{Data availability}

The data underlying this article will be shared on reasonable request to the corresponding author.





\bibliographystyle{mnras}
\bibliography{references}


\bsp	
\label{lastpage}
\end{document}